\documentclass[aps,prx,twocolumn,superscriptaddress,showpacs,floatfix,longbibliography]{revtex4-1}
\usepackage{amsmath,amssymb,graphicx}

\usepackage[utf8]{inputenc}
\usepackage[T1]{fontenc}
\usepackage{xcolor}

\IfFileExists{newtxtext.sty}
   {\usepackage{newtxtext,newtxmath}}
   {\IfFileExists{stix.sty}
      {\usepackage{stix}}
      {\IfFileExists{mathptmx.sty}
      {\usepackage{mathptmx}}{} } }

\usepackage{textcomp}

\usepackage{bm}

\IfFileExists{siunitx.sty}{\usepackage{booktabs,siunitx}}{}

\pdfoutput=1
\usepackage{color}
\definecolor{LinkColor}{rgb}{0.256,0.439,0.588}
\usepackage{hyperref}
\hypersetup{
   pdfauthor={abc},
   pdftitle={paper},
   colorlinks=true,
   citecolor=LinkColor,
   linkcolor=LinkColor,
   urlcolor=LinkColor
}

\renewcommand{\vec}[1]{\mathbf{#1}}

\usepackage{pifont}

\begin{document}

\title{Revealing Fermionic Quantum Criticality from New Monte Carlo Techniques}

\author{Xiao Yan Xu}
\affiliation{Department of Physics, Hong Kong University of Science and Technology, Clear Water Bay, Hong Kong, China}

\author{Zi Hong Liu}
\affiliation{Beijing National Laboratory for Condensed Matter Physics and Institute of Physics, Chinese Academy of Sciences, Beijing 100190, China}
\affiliation{School of Physical Sciences, University of Chinese Academy of Sciences, Beijing 100190, China}

\author{Gaopei Pan}
\affiliation{Beijing National Laboratory for Condensed Matter Physics and Institute of Physics, Chinese Academy of Sciences, Beijing 100190, China}
\affiliation{School of Physical Sciences, University of Chinese Academy of Sciences, Beijing 100190, China}

\author{Yang Qi}
\affiliation{Center for Field Theory and Particle Physics, Department of Physics, Fudan University, Shanghai 200433, China}
\affiliation{State Key Laboratory of Surface Physics, Fudan University, Shanghai 200433, China}
\affiliation{Collaborative Innovation Center of Advanced Microstructures, Nanjing 210093, China}

\author{Kai Sun}
\affiliation{Department of Physics, University of Michigan, Ann Arbor, MI 48109, USA}

\author{Zi Yang Meng}
\affiliation{Beijing National Laboratory for Condensed Matter Physics and Institute of Physics, Chinese Academy of Sciences, Beijing 100190, China}
\affiliation{Department of Physics, The University of Hong Kong, China}
\affiliation{CAS Center of Excellence in Topological Quantum Computation and School of Physical Sciences, University of Chinese Academy of Sciences, Beijing 100190, China}
\affiliation{Songshan Lake Materials Laboratory, Dongguan, Guangdong 523808, China}

\begin{abstract}
This review summarizes recent developments in the study of fermionic quantum criticality, focusing on new progress in numerical methodologies, especially quantum Monte Carlo methods, and insights that emerged from recently large-scale numerical simulations. Quantum critical phenomena in fermionic systems have attracted decades of extensive research efforts, partially lured by their exotic properties and potential technology applications and partially awaked by the profound and universal fundamental principles that govern these quantum critical systems. Due to the complex and non-perturbative nature, these systems belong to the most difficult and challenging problems in the study of modern condensed matter physics, and many important fundamental problems remain open. Recently, new developments in model design and algorithm improvements enabled unbiased large-scale numerical solutions to be achieved in the close vicinity of these quantum critical points, which paves a new pathway towards achieving controlled conclusions through combined efforts of theoretical and numerical studies, as well as possible theoretical guidance for experiments in heavy-fermion compounds, Cu-based and Fe-based superconductors, ultra-cold fermionic atomic gas, twisted graphene layers, etc., where signatures of fermionic quantum criticality exist.
\end{abstract}


\date{\today}

\maketitle

\tableofcontents

\section{Introduction}
In the past few decades, the discovery of novel quantum materials and many-body states whose unconventional properties do not fitted into traditional quantum many-body paradigms such as the Fermi liquid theory of metals and the Landau-Ginzburg-Wilson framework for phases and phase transitions, urged for new theoretical principles and insights to better understand these new quantum states and to guide the search for novel quantum materials.

In conventional materials, such as a noble metal, the net effect of interactions is limited to modify certain quantitative properties of electrons, such as the effective mass, which results in the formation of the so-called ``quasiparticle". In conventional materials, quasiplarticles at the Fermi surface has  infinite lifetime at zero temperature, which implies that these quasiparticles are effectively free from scatterings and thus behave as non-interacting particles. In the field theory description, the infinite lifetime leads to a finite quasiparticle residue, and as long as the quasiparticle residue remains finite, the non-interacting picture remains asymptotically accurate at low-energy and thus a complicated many-body quantum system is now transformed into a non-interacting system, whose physical properties can be easily understood and predicted. This approach, known as the Fermi liquid (FL) theory, has been proved to be highly effective in the study of conventional materials and it is the foundation for our understanding about many-body electronic systems.

However, in the past few decades, it is found that the Fermi-liquid theory fails in many quantum materials, such as the Cu- and Fe-based superconductors, heavy-fermion metal, and transition-metal alloys on the brink of magnetic orders~\cite{Stewart2001,Custers2003,Loehneysen2007}. These materials support the so-called `non-Fermi-liquid (nFL)" or ``strange metal" phase, in which the quasiparticle residue vanishes as the temperature is reduced to zero, instead of remaining finite as in a FL. The vanishing quasiparticle residue violates the key assumption of the FL theory, and thus results in novel quantum properties in direct contrast to FL predictions. In the studies of nFLs, one of the central focuses is systems with itinerant quantum critical points (IQCPs), which have attracted extensive research efforts for nearly half century dating back to the celebrated Hertz-Millis-Moriya (HMM) framework~\cite{Hertz1976,Millis1993,Moriya1985}. In the study of correlated quantum materials, quantum criticality in itinerant electron systems is of great importance and interests~\cite{Hertz1976,Millis1993,Moriya1985,Stewart2001,Chubukov2004,Belitz2005,Loehneysen2007,Chubukov2009}, and it plays an important role in the understanding of anomalous transport, strange metal and nFL behaviors~\cite{Metzner2003,Senthil2008,Holder2015,Metlitski2015,Xu2017} in various quantum materials, such as heavy-fermion materials~\cite{Custers2003,Steppke2013}, Cu- and Fe-based high-temperature superconductors~\cite{ZhangWenLiang2016,LiuZhaoYu2016,Gu2017,ChunGuangWang2018}, the recently discovered  pressure-driven quantum critical point (QCP) between magnetic order and superconductivity in transition-metal monopnictides, CrAs~\cite{Wu2014}, MnP~\cite{Cheng2015}, CrAs$_{1-x}$P$_x$~\cite{JGCheng2018} and other Cr/Mn-3d electron systems~\cite{Cheng2017} and the more recent discoveries in twisted angle graphene heterostructures~\cite{cao2018correlated,YuanCao2019,ChengShen2019}. However, after decades of extensive efforts~\cite{Hertz1976,Millis1993,Moriya1985,Stewart2001,Metzner2003,Abanov2003,Abanov2004,Chubukov2004,Belitz2005,Loehneysen2007,
Chubukov2009,Metlitski2010a,Metlitski2010b,Sur2015,Sur2016,Schlief2017,Lunts2017,SSLee2018,Schlief2018}, itinerant quantum criticality still remains among the most challenging subjects in condensed matter physics due to its nonperturbative nature, and many fundamental questions and puzzles still remain open.

It is worthwhile to point out that similar to classical critical phenomena and thermal phase transitions, the theoretical understanding of quantum criticality and electronic quantum criticality requires combined efforts from both analytic theory and large-scale numerical simulations. In the past few decades, a lot of great progress have been made on the theory side~\cite{Hertz1976,Millis1993,Moriya1985,Stewart2001,Metzner2003,Abanov2003,Abanov2004,Chubukov2004,Belitz2005,Loehneysen2007,
Chubukov2009,Metlitski2010a,Metlitski2010b,Sur2015,Sur2016,Schlief2017,Lunts2017,SSLee2018,Schlief2018}. However, numerically testing and verifying these predictions remain a highly challenging task for IQCPs in 2D or higher dimensions, mainly due to limitations in all available numerical techniques. The absence of reliable and unbiased numerical verification makes it difficulty to verify key assumptions adopted in theoretical studies and thus making it highly challenging to refine our theoretical knowledge about IQCPs.

Recently, thanks to major new progress in quantum Monte Carlo (QMC)  techniques, accurate and unbiased numerical solutions becomes achievable for various IQCPs.
It is found that designer models of fermions coupled to critical bosonic fields offer a pathway to access a wide variety of IQPCs, and in the same time avoid the notorious sign-problem large-scale QMC simulations possible~\cite{Berg2012}. Utilizing this approach, a variety of fermionic quantum critical points are studied, with critical/soft fluctuations such as antiferromagnetic, nematic or phononic fluctuations~\cite{Berg2012,Schattner2016,Schattner2015b,Li2016925}.  In the study of unconventional superconductors, such as Cu- or Fe- based high temperature superconductors, these soft low-energy fluctuations have been experimentally observed. However, whether these fluctuations are the main driving force for the formation of high temperature superconductivity remains an important open questions, and these QMC simulations are expected to shed new lights on this long-standing question. Remarkably, all these QMC simulations shares one common observation: as one turns on the coupling between fermions and critical fluctuations, a superconducting dome always arises and covers the QCP regardless of  the symmetry breaking patterns, which suggests that there might exist a deep and fundamental connection between high temperature superconductivity and these quantum critical fluctuations.

In this review, we focus on a different aspect of IQCPs. Instead of exploring superconducting domes induced by IQCPs, our objective is to understand the IQCPs themselves, i.e. to identify the critical theory and  its universal scaling behavior and to explore the non-Fermi liquids induced by the critical flucuations. For this purpose, we need to access the close vicinity of a QCP, which requires suppressing the superconducting dome to expose the QCP.  Experimentally, superconductivity can be in principle suppressed by a strong magnetic field, but for QMC studies, this approach is not the optimum pathway, because handling gauge fields is numerical challenging. Instead, in QMC, we found that a more convenient approach is to use designer Hamiltonians, especially using bilayer models that contain two identical copies of fermions~\cite{Xu2017}. Although the mechanism is not yet fully understand, in all bilayer models that we have examined, this setup suppresses the superconducting dome down to extremely low temperature unaccessible to QMC simulations. In addition to suppressing superconductivity, to obtain accurate critical phenomena and scaling exponents, large system sizes, in both spatial and temporal directions, are necessary, which requires new numerical techniques to access larger systems and lower temperature.

Along this line of efforts, new designer models and QMC simulation techniques are developed to examine in detail the properties of IQCPs. For example, to study strange metal behaviors near the ferromagnetic quantum phase transition, we developed the self-learning quantum Monte Carlo method~\cite{JWLiu2017SLMC,XYXu2017SLMC} to solve this problem and observed clear signatures of non-Fermi-liquid behaviors near the IQCP~\cite{Xu2017}. More remarkably, quantum critical analysis indicates it is a new critical point with an anomalous scaling exponent~\cite{Xu2017}. These results not only sharpen the theoretical understanding about IQCP, but also offer important theoretical guidance to experimental investigations on various strongly correlated metals. The similar technique is also applicable for antiferromagnetic QCPs, via coupling antiferromagnetic fluctuations with itinerant fermions~\cite{Berg2012,ZiHongLiuTri2018,ZiHongLiuEMUS2018,ZiHongLiuSqu2018}, which can be used to investigate the fundamental phenomena such as antiferromagnetic fluctuations mediated charge density wave, psesudo-gap phase and unconventional superconductivity. All these questions can now be hopefully addressed with unbiased numerical simulations. Moreover, with this technique, one can also study the effect of compact gauge fields coupled to a Fermi surface, such that the interdisciplinary research between condensed matter and high energy physics on the subject of quantum spin liquid with emergent gauge field and
the confined and deconfined state of matter can be explored~\cite{XiaoYanXu2018U1}. All these fast developments, by and large, are calling for a topical review such as this in front of the readers, to summarize the key developments and point out future directions.

The content of the review is based on several recent numerical works. 
The discussion about ferromagnetic itinerant QCP and the associated nFL quantum critical region (QCR), is mainly based on Ref.~\cite{Xu2017}. The discussion on antiferromagnetic IQCP, in realization of 2d triangle and square lattices and their similarity and difference, are mainly based on Refs.~\cite{ZiHongLiuTri2018,ZiHongLiuSqu2018}. The developments in the QMC methodologies, including basic outline of the determinantal QMC (DQMC) for fermion coupled to critical boson problems are based on Refs.~\cite{xu_thesis,ZHLiuTutorial}, and the framework of self-learning Monte Carlo (SLMC in short) methods are based on the Refs.~\cite{JWLiu2017SLMC,XYXu2017SLMC,ChuangChen2018a} as well as the momentum-space based elective momentum ultra-size quantum Monte Carlo method (EMUS in short) are based on the Refs.~\cite{ZiHongLiuEMUS2018,ZiHongLiuSqu2018}. The other recent developments, including the application of SLMC on electron-phonon coupled Holstein-type of problems~\cite{ChuangChen2018a,ChuangChen2018b} and the question of U(1) gauge fields coupled to matter fields in the form of Dirac fermions and the deconfinement-to-confinement transition discoveried in the such model simulations~\cite{XiaoYanXu2018U1}, are discussed as well.

Because this review mainly focuses on QMC studies, we will also provide a pedagoical QMC code package developed over the past few years by some of us, which contains the lattice construction, local update of conventional DQMC and SLMC steps with learning of effective model and the cumulative update scheme afterwards, as well as the EMUS construction of the momentum patches. The package can be found in the GitHub repository~\cite{Xudqmc_demo} and the corresponding manual can be found in the Ref.~\cite{ZHLiuTutorial}. Readers who are interested in understanding the numerical details of these new Monte Carlo techniques and would like to reproduce our results and carry out further investigations, are strongly encouraged to download, install and run the package and modify it to his/her purposes. And the developers are more than willing to offer help if questions/problems/comments arise.

\section{Theoretical setting}
The description of quantum phase transitions and quantum criticality in metallic systems was pioneered by the Hertz-Millis-Moriya theory~\cite{Hertz1976,Millis1993,Moriya1985}.
However, it was soon found that Hertz-Millis-Moriya (HMM) formalism may not be adequate, and for a number of important situations, higher order terms ignored in the Hertz-Millis-Moriya are found to result in unstable RG flows away from the Hertz-Millis-Moriya fixed point, such as the ferromagentic, antiferromagnetic and nematic quantum critical points in (2+1)D. And more advanced theoretical frameworks, aiming at capturing these quantum critical points and more importantly the non-Fermi-liquid behavior associated with them, are developed~\cite{Metzner2003,Senthil2008,Metzner2003,Abanov2003,Abanov2004,Chubukov2004,Belitz2005,
Chubukov2009,Metlitski2010a,Metlitski2010b,Dalidovich2013,SSLee2009,Sur2016,Schlief2017,SSLee2018,Schlief2018} over the years. Although this review mainly focuses on the numerical end, we will briefly introduce the theoretical setting and describes its fundamental challenges in simple terms.

\subsection{Model}
For strongly-correlated systems, although theoretical descriptions often start from fermions with direct fermion-fermion interactions (e.g. four-fermion interactions), to describe a quantum phase transition as often used in the Hertz-Millis-Moriya approach~\cite{Hertz1976,Millis1993,Moriya1985}, the most efficient way is to use an effective model where directly fermion-fermion interactions are replaced by certain boson-mediated interactions. These bosons can be thought of as the auxiliary field in the Hubbard-Stratonovich transformation. Here, we choose the auxiliary filed to share the same symmetry as the order parameter of the phase transition, i.e. an order-parameter field. Thus, the expectation value of the boson operator serves as the order parameter and these bosonic fields describes soft critical fluctuations in the vicinity of the QCP. It is worthwhile to emphasize that this effective model don't fully replicate the details of the original model with fermion-fermion interactions. In particular, contributions from other quantum fluctuations beyond the order parameter field is ignored in the effective theory. However, near the QCP, universal properties, which are independent of microscopic details, are expected to survive.

Interestingly, this idea of using an effective model also plays an important role in QMC studies. For QMC simulations, directly simulating four-fermion interactions often turns to be challenging due to the arising of the sign problem. However, the effective model used in the Hertz-Millis-Moriya approach (with boson-mediated interactions) turns out to be much more QMC friendly and the sign problem can often be avoided. Thus, although a generic and efficient way to avoid the sign problem is still absent for the Hubbard or Hubbard-like models, the sign problem can be easily avoided in many of these boson-mediated effective model. As mentioned above, these effective models are precisely the setup utilized in most of the theoretical studies and they are expected to capture universal properties at IQCPs. Thus, these effective models with boson-mediated interactions opens a pathway to study these IQCPs via large scale numerical simulations.

To demonstrate the effective theory used in the study of fermionic quantum criticality, here we show one example based on an Ising ferromagnetic transition in continuous space, but the same construction generically applies to other quantum phase transitions as well as lattice systems. Consider the following action.
\begin{equation}
 S=S_{\text{fermion}}+S_{\text{boson}}+S_{\text{coupling}}
\label{eq:model}
\end{equation}
Here $S_{\text{fermion}}$ is the action of a non-interacting fermion gas
\begin{align}
 S_{\text{fermion}} =\int d\mathbf{r} dt \sum_{\alpha} \left( c_\alpha^\dagger i\partial_t c_\alpha- \frac{\hbar^2}{2 m_e}\nabla c_\alpha^\dagger \nabla c_\alpha \right),
\end{align}
where $c_\alpha^\dagger$ and $c_\alpha$ are the fermion creation and annihilation operators with spin index $\alpha=\uparrow$ or $\downarrow$. $m_e$ and $\hbar$ are the effective
mass of the fermion and the Planck constant respectively.
The second term $S_{\text{boson}}$ is the action of a bosonic field. This boson field respects the same symmetry as the order parameter of quantum phase transition and this action includes all symmetry allowed terms of the boson field, as well as symmetry allowed quantum dynamics. For an Ising transition, we can use the $\phi^4$-theory here
\begin{align}
 S_{\text{boson}} =\frac{1}{2}\int d\mathbf{r} dt \left[ (\partial_t \phi)^2- (\nabla \phi)^2 - m  \phi^2 - \frac{u}{2}  \phi^4 \right].
\end{align}
Without fermions, this action describe a bosonic quantum phase transition with the same symmetry breaking pattern. In the example here, a bosonic QCP arises as we decrease the value $m$ to negative. The last term $S_{\text{coupling}}$ is the coupling term between fermions and bosons. The boson field is coupled with a fermion bilinear, which respects the same symmetry as the order parameter. In the example here, the fermion bilinear is z-component of the fermion spin $s_z= c_\alpha^\dagger \sigma^z_{\alpha,\beta} c_\beta$ where $\sigma^z$ is the Pauli matrix and the action is
\begin{align}
 S_{\text{coupling}} =g \int d\mathbf{r} dt  \; \phi \; s_z,
\end{align}
where $g$ is the coupling constant.

In this model, the interactions between fermions are mediated by the bosonic field $\phi$. And in return, fermionic fluctuations also renormalize the bosonic part of the action.
At $g=0$, when the bosons and fermions decouple, the susceptibility of the bosonic order parameter field (at the tree level) is
\begin{equation}
\chi_{0}(\mathbf{q},\omega)=\frac{1}{\omega^{2}-q^2-\xi_c^{-2}},
\label{eq:bareboson}
\end{equation} 
where $\xi_c$ is correlation length of bosoinc field, which diverges at the critical point. As we turn on the boson-fermion coupling ($g\ne 0$), a phase transition is still expected as we reduce
the value of $m$, although the critical threshold $m_c$ is now renormalized. This construction demonstrate a generic approach to realize and study fermionc quantum phase transitions.

\begin{figure*}[htp!]
	\includegraphics[width=0.8\textwidth]{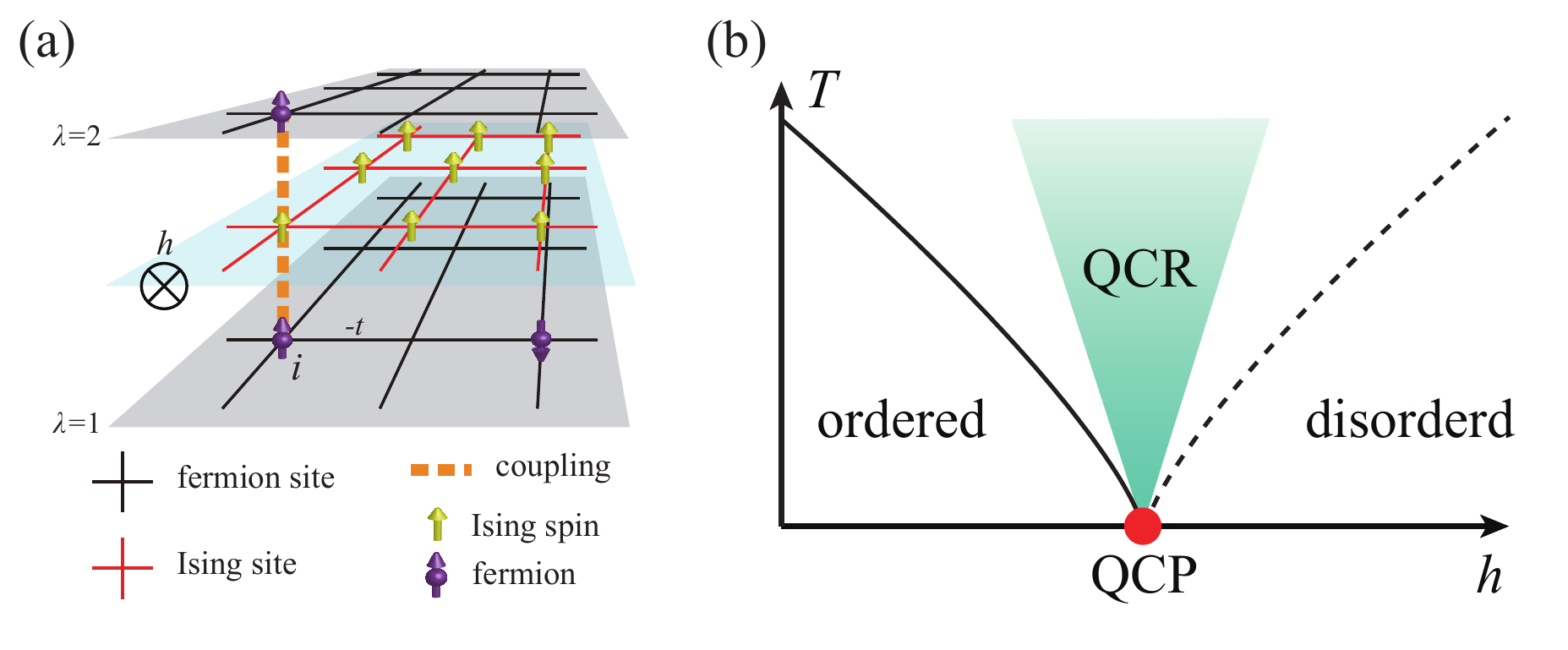}
	\caption{(a) Illustration of our designer model in Eq.~\eqref{eq:specificformula}. Fermions reside on two of the	layers ($\lambda$ = 1,2) with intra-layer nearest-neighbor hopping $t$. The middle layer is composed of Ising spins $s_{i}^{z}$ , subject to nearest-neighbor antiferromagnetic Ising coupling $J$ and a transverse magnetic field $h$. Between the layers, an onsite Ising coupling is introduced between fermion and Ising spins ($\xi$). (b) Illustration of phase diagram of the designer model. The ordered phase spontaneously breaks the Ising symmetry (defined in the mean text). In the ordered phase (below the black solid phase boundary) or disordered region (below the black dashed line -- the crossover boundary), Ising spins largely align along the $z$ or $x$ axes, respectively, and thus quantum fluctuations are limited and under control. In the quantum critical region (QCR marked by the green area), strong critical fluctuations arise, driving the system into the nonperturbative strong-coupling regime. Theoretical treatment for this regime is highly challenging. For this QCP, unbiased quantum Monte Carlo simulations reveal a new fermionic quantum criticality with exponents deviates from all existing theoretical predictions. Panel (a) is adapted from Ref.~\cite{Xu2017}.}
	\label{fig:fig1}
\end{figure*}

\subsection{the Hertz-Millis-Moriya theory and beyond}
\label{sec:HMM}
As boson mediates interactions between fermions, the fermion also mediates nontrivial couplings among the bosons. This can be seen by integrating out the fermionic degrees of freedome in the action of Eq.~\eqref{eq:model}, which leads to a bosonic effective theory for a fermionic QCP. In the Hertz-Millis-Moriya theory, this is done within the framework of the random phase approximation (RPA). Namely, only the so-called bubble diagram is included while higher order contributions are ignored. 

With this assumption, the boson susceptibility receives significant renormalization
\begin{align}
\chi(\mathbf{q},\omega)=\frac{\chi_0}{a_0 i |\omega|/q+a_1 \omega^{2}- a_2 (q^2+\xi_c^{-2})},
\end{align}
where $\chi_0$, $a_1$ and $a_2$ are renormalized parameters, whose values goes to unity in the $g \rightarrow 0$ limit where the susceptibility without fermions in the form of Eq.~\eqref{eq:bareboson} is recovered. Aside from renormalizing parameters at the quantitative level, the most significant contribution from the fermions lies in the new quantum dynamics it introduced to $\phi$, i.e. the term $i |\omega|/q$, known as Landau damping. The coefficient of this term, $a_0$, depends on the coupling strength $g$. It is finite for $g\ne 0$ and vanishes in the $g \to 0$ limit. For finite $g$, in the low-energy limit (small $\omega$), this Landau-damping term dominates over the old $\omega^{2}$ term. In the language of renormalization group, this term is more relevant and thus making the old $\omega^2$ term subleading and negligible. After dropping the sub-leading $\omega^2$ term, a dimension counting indicates that this quantum critical point shall have the dynamic critical exponent $z=3$~\cite{Hertz1976}. This Landau damping term and the dynamic critical exponent $z=3$ are generically expected for any phase transition that preserves the (lattice) translational symmetry (e.g. Ising feromagnetic or nematic). For QCPs at which the (lattice) translational symmetry is spontaneously broken (e.g. charge- or spin- density wave states), the Landu damping term takes a slightly different form $\propto i |\omega|$, which often results in $z=2$~\cite{Hertz1976}.

As a result, with in the Hertz-Millis-Moriya approximation, the fermions push the effective dimension of a fermionic QCP to $d+z$. For $d\ge 2$, no matter the value of $z$ is $2$ or $3$, $d+z$ will be above or equal to the upper critical dimension $4$, and thus mean-field exponents is expected regardless of microscopic details, which is a key prediction of the Hertz-Millis-Moriya theory. This prediction is highly generic. Although RG flows may lead to nontrivial temperature dependence beyond the RPA approximation~\cite{Millis1993}, at $T=0$, the RPA prediction and mean-field exponent is always expected within this framework.

However, it is worthwhile to highlight that this strong and universal prediction is based on one important assumption: as we integrate out the fermions, the HMM approach assumes that only leading order one-loop diagram (i.e. the bubble diagram) needs to be included. This assumption can be justified in a Fermi liquid phase. However, at a fermionic QCP, non-Fermi-liquid behaviors often emerge and it is highly unclear whether this  one-loop approximation remains valid or not. In particular, as higher order diagrams are taken into account, latter studies indicate that certain fermionic QCPs exhibit a highly nontrivial scaling behavior, with scaling exponent significantly different from mean-field expectations~\cite{Abanov2003,Metlitski2010a,Metlitski2010b,SSLee2009,SSLee2018}. In particular, for antiferromagnetic critical point, these anomalous dimensions are computed within the  framework of $1/N$ expansion, where $N$ is the number of the fermion hotspots, which will be defined below. Anomalous dimensions are observed even in the first order correction ($\propto 1/N$)~\cite{Abanov2003}, and later, more careful analysis indicates that the dynamic critical exponent also receives an anomalous dimensions~\cite{Metlitski2010b}. And self-consistent theory is developed to predict the RG flow of the  dynamic critical exponent~\cite{SSLee2018}. Similar breakdown of HMM scaling was also predicted for some other fermionic QCPs~\cite{Metlitski2010a}.

Although this non-mean-field exponents has been predicted  for years, in contrast to the long efforts and rapid progress in analytic theory, progress on the numerical front had a slower pace and no signature of anomalous dimension has been achieved due to system size and other limitations, until very recently. Below, we will summarize some of the latest progress along this line and in particular, we focuses on two questions: (1) whether it is possible for numerical techniques to achieve high enough resolution to distinguish mean-field and non-mean-field exponents and (2) whether we can determine numerically which fermionic QCPs have mean-field exponents (i.e. the Hertz-Millis-Moriya theory  remains valid) and which fermionic QCPs exhibit the breakdown of the Hertz-Millis-Moriya theory.

\section{Numerical methodologies}
Before showing the numerical results that answer the second question above, in this section we address the first question on the development of numerical methodologies.

\subsection{Designer-Hamiltonians}
To put the problem in Eq.~\eqref{eq:model} on a lattice and solve with quantum Monte Carlo, one would write down the following microscopic Hamiltonian,
\begin{equation}
H=H_{\text{fermion}}+H_{\text{boson}}+H_{\text{coupling}},
\label{eq:generalformula}
\end{equation}
with
\begin{eqnarray}
H_{\text{fermion}} & = & -t_{ij} \sum_{ij,\alpha}(c_{i\alpha}^{\dagger}c_{j\alpha}+h.c.)-\mu \sum_{i}n_{i}\\
H_{\text{boson}} & = & J \sum_{ij}s_{i}^{z}s_{j}^{z}-h\sum_{i}s_{i}^{x}\\
H_{\text{coupling}} & = & -\xi \sum_{ijk,\alpha\beta}c_{i\alpha}^{\dagger}s_{k}^{z}c_{j\beta}
\end{eqnarray}
where $H_{\text{fermion}}$ is the free fermion Hamiltonian term; $H_{\text{boson}}$
is the bosonic Hamiltonian realized in the form of a transverse-field quantum Ising model and $H_{\text{coupling}}$ is the couple interaction
term between the Ising spin and the fermion. $i$, $j$ and $k$ are the lattice site indices and $\alpha$ contains the set of the orbital and spin and other internal freedom. The free fermion Hamiltonian $H_{\text{fermion}}$ is the hopping term in real space,
after Fourier transformation, we can obtain the energy band in momentum space
\begin{eqnarray}
H_{\text{fermion}}&=&-t_{ij}\sum_{ij,\alpha}(c_{i\alpha}^{\dagger}c_{j\alpha}+h.c.)-\mu \sum_{i}n_{i} \nonumber\\
&=&\sum_{k}(\epsilon(k)-\mu)c_{k}^{\dagger}c_{k}.
\end{eqnarray}
The Ising spin term $H_{\text{boson}}$ makes use of the tuning parameter $h_{x}/J$ to obtian an quantum phase transition between the magnetic ordered phase and the disordered paramagnetic phase at zero temperature. The coupling term $H_{\text{coupling}}$ plays a vital role in introducing the critical fluctuations into the fermions and vice versa. The fermion-spin coupling will move the location of the quantum critical point. More importantly, because the fermion-spin coupling is a relevant perturbation in RG, it will change the critical exponents and transfers the quantum critical points into a different universality class. 

To be more specific, let us consider a 2D system with three layers of square lattices, including two fermion layers with index $\lambda$ and spin index $\sigma$, and one Ising-spin layer lattice inserted in between the two fermion lattice layers, as schematically shown in Fig.~\ref{fig:fig1}. At each lattice site, the Ising spin couples with fermion spins. Such type of specific designer-model, as studied in Refs.~\cite{Xu2017,ZiHongLiuTri2018,ZiHongLiuEMUS2018,ZiHongLiuSqu2018}, can be written as
\begin{eqnarray}
H_{\text{fermion}} & = & -t \sum_{\left\langle i,j\right\rangle \sigma\lambda}(c_{i\lambda\sigma}^{\dagger}c_{j\lambda\sigma}+h.c.)-\mu \sum_{i\sigma\lambda}n_{i\lambda\sigma} \nonumber\\
H_{\text{boson}} & = & J\sum_{\left\langle i,j\right\rangle }s_{i}^{z}s_{j}^{z}-h \sum_{i}s_{i}^{x}\nonumber\\
H_{\text{coupling}} & = & -\xi \sum_{i}s_{i}^{z}(c_{i}^{\dagger}\tau_{0}\sigma_{z}c_{i})=-\xi \sum_{i}s_{i}^{z}(\sigma_{i1}^{z}+\sigma_{i2}^{z}).\nonumber\\
\label{eq:specificformula}
\end{eqnarray}
In Eq.~\eqref{eq:specificformula} the basis of the fermion Hilbert space is $\tau\otimes\sigma\otimes i$. The coupling between fermion and boson is onsite in $H_{\text{coupling}}$. $c_{i}^{\dagger}$ is a four component spinor, $c_{i}^{\dagger}=\begin{bmatrix}c_{i1\uparrow}^{\dagger} & c_{i1\downarrow}^{\dagger} & c_{i2\uparrow}^{\dagger} & c_{i2\downarrow}^{\dagger}\end{bmatrix}$,
and $\tau_{0}$ is a $2\times 2$ identity matrix in layer/orbital space, $\sigma_{z}$ is the third Pauli matrix in spin space, respectively. $\sigma_{i\lambda}^{z}=\frac{1}{2}(n_{i\lambda\uparrow}-n_{i\lambda\downarrow})$
is the $z$ component of the fermion spin at orbit $\lambda$ on site
$i$. This system preserves a $Z_{2}$  Ising symmetry corresponding to the Ising spin flipping $s_{i}^{z}\rightarrow-s_{i}^{z}$, which will be spontaneously broken in the ordered phase.

As shown in the Fig.~\ref{fig:fig1} (b), the Ising spins develop long-range Ising order in the ordered phase. Deep in the ordered phase, (small $h$ and low $T$), quantum fluctuations of the Ising fields diminish and thus, the Ising spins can be treated as a static and average background, i.e. a mean field. Through the coupling term $H_{\text{coupling}}$, this mean field formed by Ising spins induce the same symmetry breaking pattern to the fermions, and this is how a fermionic phase transition is achieved in these type of models. This construction is highly flexible and generic. Any type of symmetry breaking phase be achieved, as long as proper boson (Ising) fields that break the desired symmetry is utilized. In systems with itinerant fermions, quantum phase transition can be largely classified into two categories depending on whether the ordered phase preserves  (lattice) translational symmetry or spontaneously breaks it. Both these two types of quantum phase transition has been studied recently using the QMC approach described here. Examples in the first category includes ferromagnetic~\cite{Xu2017} and nematic~\cite{Schattner2016} phase transitions in metals, while antiferromagnetic quantum phase transition (with itinerant fermions)~\cite{Gerlach2017,ZiHongLiuSqu2018,ZiHongLiuTri2018} belongs to the second category. As will be shown in latter sections, these symmetry-breaking metallic phases are still FLs with well-defined quasiparticles, after certain Fermi surface reconstructions, e.g. splitted, distorted or folded Fermi surfaces due to the symmetry-breaking and/or the induced Brillouin zone folding.

The real surprise of these designer models [Eq.~\eqref{eq:specificformula}] arises near the quantum critical point in the quantum critical region (QCR), which is also the most challenging regime for theoretical treatments. Inside QCR, the critical bosons cannot be viewed as a mean field any more, and their strong quantum critical fluctuations mediate intensive effective interactions between fermions, which pushes the system into the strongly-correlated regime and is also the origin of non-Fermi-liquid behaviors in these systems.  
These strong interactions make theoretical description of such a quantum critical regime highly challenging, and exactly solution is not expected. In fact, almost the entire efforts in the past four or five decades, starting from HMM as explained in Sec.~\ref{sec:HMM}, focuses on employing or inventing new and advanced theoretical (renormalization group) machinery, to extract the effective low-energy interactions in these systems. Such effective fermionic and bosonic interactions are not only non-local in space and also non-local in time, rendering the entire problem unperturbative and consequently prohibiting controlled analytical solutions. It is such kind of theoretical challenge, motivated the development of unbiased quantum Monte Carlo methods, to tackle these problems and to bring new insights into our understanding of fermionic quantum criticality. Along the way, better numerical methodologies are inspired and invented with guidance from field theory analysis, from which new results that are bringing us coherent understanding between both numerical and theoretical perspectives, are revealed. In the two subsections followed, we will discuss the quantum Monte Carlo methodologies used and developed during this interesting process.

\subsection{Determinantal quantum Monte Carlo}

To solve the model in Eq.~\eqref{eq:specificformula}, one makes use of the determinant quantum Monte Carlo method. For the purpose of this review, the basic literatures that introduce the method are  Refs.~\cite{AssaadEvertz2008,xu_thesis,ZHLiuTutorial}. DQMC plays a vital role in the study of correlated electron models. As an unbiased method, DQMC can solve  sign-problem-free strongly-correlated fermion models with large-scale parallelization to access the thermodynamic limit. Historically, it was first introduced to condensed matter physics systems from high-energy physics community by Blankenbecler, Scalapino and Sugar, so the method is also referred to as BSS method~\cite{BSS}. DQMC
is based on the canonical ensemble within a path integral formalism. Starting from $d$ dimensional quantum grand canonical ensemble $Z=\mathrm{Tr}\{e^{-\beta H}\}$,
and use the path integral formalism to tranform the partition function
into the $(d+1)$ dimensional classical partition. In the $(d+1)$
dimensional classical partition, the partition function $Z$ is represented as the sum over weights of configuration, which can be seen as the probability of configurations.

In the past, DQMC is mostly used to solve Hubbard or Hubbard-like models with explicit four-fermion interactions~\cite{AssaadEvertz2008,Varney2009,Meng2010,YYHe2016}. And to be able to construct the path-integral formalism of the partition function, one needs to introduce auxiliary field to decouple the four-fermion interaction (usually refered as discrete Hubbard-Stratonovich (HS) transformation)~\cite{Hirsch1983,Hirsch1985}. After the decoupling, bosonic HS fields will emerge along with fermion bilinears and it is the space-time configuration of the HS fields that one could use to compute the fermion determinant and the determinant plays the role to the configurational weight, upon which the Markov-chain of Monte Carlo is built.

However, in the new types of problem as shown in Eq.~\eqref{eq:specificformula}, the spin-fermion model already acquires the decoupled form, or in another word, the boson fields in Eq.~\eqref{eq:specificformula} already couples to the fermion bilinears, so we do not need to perform the HS transformation. The configurational space in such cases are spanned by the space-time configurations of the boson fields, and for each boson configuration, one can trace out the fermions by evaluating the fermion determinant. Below we first discuss how to construct the bosonic weight of the bare transverse field Ising model $H_{\text{boson}}$.

It is easily to see that from  
\begin{equation}
e^{\Delta\tau hs_{i}^{x}}=\cosh(\Delta\tau h)\boldsymbol{1}+\sinh(\Delta\tau h)s_{i}^{x},
\end{equation}
one obtains
\begin{equation}
\left\langle S_{z}^{\prime}\right|e^{\Delta\tau h s_{i}^{x}}\left|S_{z}\right\rangle =\Lambda e^{\gamma s_{z}^{\prime}s_{z}}.
\end{equation}
For the Ising field in $H_{\text{boson}}$ $S_{z}=\pm1$, so we can get 
\begin{eqnarray}
\langle S_{z} |e^{\Delta\tau h s_{i}^{x}}|S_{z}\rangle &=&\cosh(\Delta\tau h)=\Lambda e^{\gamma} \nonumber\\
\langle -S_{z} |e^{\Delta\tau h s_{i}^{x}}|S_{z}\rangle &=& \sinh(\Delta\tau h)=\Lambda e^{-\gamma}.
\end{eqnarray}
then $\Lambda$ and $\gamma$ can be evaluated as 
\begin{eqnarray}
\gamma &=& -\frac{1}{2}\tanh(\Delta\tau h)\nonumber\\
\Lambda^{2} &=& \sinh(\Delta\tau h)\cosh(\Delta\tau h).
\end{eqnarray}
Defining $|S_{z}^{N}\rangle$ as the product state of Ising spin at each site in time slice $\tau=N\Delta\tau$, $\beta=M\Delta\tau$ and impose the periodic condition in time direction $|S_{z}^{M}\rangle =|S_{z}^{1}\rangle$. We can rewrite the bare transverse field Ising model partition function
\begin{widetext}
\begin{eqnarray}
Z & = & \text{Tr}\{e^{-\beta H_{\text{boson}}}\} \nonumber\\
  & = & \sum_{\{S_{z}\}}\langle S_{z}^{M}|e^{-\Delta\tau H_{\text{boson}}}|S_{z}^{M-1}\rangle\langle S_{z}^{M-1}|e^{-\Delta\tau H_{\text{boson}}}|S_{z}^{M-2}\rangle\langle S_{z}^{M-2}|\cdots|S_{z}^{2}\rangle\langle S_{z}^{2}|e^{-\Delta\tau H_{\text{boson}}}|S_{z}^{1}\rangle + O(\Delta\tau^{2})\nonumber\\ 
 & = & \sum_{\{S_{z}\}}\langle S_{z}^{M}|e^{-\Delta\tau J\sum_{\langle i,j\rangle}s_{i}^{z}s_{j}^{z}}e^{\Delta\tau h \sum_{i}s_{i}^{x}}|S_{z}^{M-1}\rangle\cdots\langle S_{z}^{2}|e^{-\Delta\tau J\sum_{\langle i,j\rangle}s_{i}^{z}s_{j}^{z}}e^{\Delta\tau h \sum_{i}s_{i}^{x}}|S_{z}^{1}\rangle +O(\Delta\tau^{2}) \nonumber\\
&=& (\prod_{\tau}\prod_{\langle i,j \rangle} e^{-\Delta\tau J\sum_{\langle i,j \rangle}s_{i,\tau}^{z}s_{j,\tau}^{z}})(\prod_{i}\prod_{\langle\tau,\tau'\rangle}\Lambda e^{\gamma\sum_{i}s_{i,\tau}^{z}s_{i,\tau'}^{z}})+O(\Delta\tau^{2})
\label{eq:bareIsing}
\end{eqnarray}
\end{widetext}
where $S_{i,\tau}^{z}=\pm1$ is the Ising field lives in $d+1$ dimensional
space after path integral. With the partition function of $H_{\text{boson}}$ obtained, one can readily write down the partition function for the entire system of Eq.~\eqref{eq:specificformula} since here there is no four-fermion interaction terms, as 
\begin{widetext}
\begin{eqnarray}
Z & = & \text{Tr}\{e^{-\beta H}\} \nonumber\\
  & = & \text{Tr}\{(e^{-\Delta\tau H_{\text{fermion}}}e^{-\Delta\tau H_{\text{boson}}}e^{-\Delta\tau H_{\text{\text{coupling}}}})^{M}\}+O(\Delta\tau^{2})\nonumber\\
 & = & \text{Tr}_{f}\{\sum_{\{S_{z}\}}\langle S_{z}^{M}|e^{-\Delta\tau H_{\text{fermion}}}e^{-\Delta\tau H_{\text{boson}}}e^{-\Delta\tau H_{\text{coupling}}}|S_{z}^{M-1}\rangle\cdots\langle S_{z}^{2}|e^{-\Delta\tau H_{\text{fermion}}}e^{-\Delta\tau H_{\text{boson}}}e^{-\Delta\tau H_{\text{coupling}}}|S_{z}^{1}\rangle\}+O(\Delta\tau^{2})\nonumber\\
 & = & \sum_{\{S_{z}\}}(\prod_{\tau}\prod_{\langle i,j\rangle }e^{-\Delta\tau J\sum_{\langle i,j\rangle}s_{i,\tau}^{z}s_{j,\tau}^{z}})(\prod_{i}\prod_{\langle \tau,\tau'\rangle}\Lambda e^{\gamma\sum_{i}s_{i,\tau}^{z}s_{i,\tau'}^{z}})\text{Tr}_{f}\{\prod_{\tau}e^{-\Delta\tau H_{\text{fermion}}}e^{-\Delta\tau H_{\text{coupling}}(\{s_{\tau}^{z}\})}\}+O(\Delta\tau^{2})\nonumber\\
 &=& \sum_{\{S_{z}\}}\mathcal{W_{C}}^{\text{boson}}{\text{Tr}}_{f}\{\prod_{\tau}(e^{-\Delta\tau c^{\dagger}Tc}\prod_{i}e^{c^{\dagger}V\{s_{i,\tau}^{z}\}c})\}+O(\Delta\tau^{2})\nonumber\\
 & = & \sum_{\{S_{z}\}}\mathcal{W_{C}}^{\text{boson}}\det[\boldsymbol{1}+\prod_{\tau}(e^{-\Delta\tau T}\prod_{i}e^{V\{s_{i,\tau}^{z}\}})]+O(\Delta\tau^{2})\nonumber\\
 &=& \sum_{\{S_{z}\}}\mathcal{W_{C}}^{\text{boson}}\det[\boldsymbol{1}+B(\beta,0)]+O(\Delta\tau^{2})\nonumber\\
 &=& \sum_{\{S_{z}\}}\mathcal{W_{C}}^{\text{boson}}\mathcal{W_{C}}^{\text{fermion}}+O(\Delta\tau^{2})
\label{eq:spin-fermion}
\end{eqnarray}
\end{widetext}
where 
\begin{align}
\mathcal{W_{C}}^{\text{boson}}=& \sum_{\{S_{z}\}}(\prod_{\tau}\prod_{\langle i,j\rangle}e^{-\Delta\tau J\sum_{\langle i,j\rangle}s_{i,\tau}^{z}s_{j,\tau}^{z}}) \nonumber \\ 
 & \ \ \ \ \ \ \  (\prod_{i}\prod_{\langle \tau,\tau'\rangle}\Lambda e^{\gamma\sum_{i}s_{i,\tau}^{z}s_{i,\tau'}^{z}})
\end{align}
is the weight of bare bosonic part in Eq.~\eqref{eq:bareIsing}, and
\begin{equation}
\mathcal{W_{C}}^{\text{fermion}}=\det[\boldsymbol{1}+\prod_{\tau}(e^{-\Delta\tau T}\prod_{i}e^{V\{s_{i,\tau}^{z}\}})]
\end{equation}
is the Femion weight which depends on the space and imaginary time
distribution of the Ising field $\{S^{z}_{i,\tau}\}$. Eq.~\eqref{eq:spin-fermion} can be sampled with Markov process according to the local update of $\{S_{i,\tau}^z\}$ spins. And with the DQMC update scheme developed over the years, such evaluation of the ratio of the fermion determinant, with propagation of the $N\times N$ matrices from $0$ to $\beta$~\cite{AssaadEvertz2008}, acquires the computational complexity scaling with the system parameters as $O(\beta N^{3})$ in the ideal case, if the autocorrelation between different configurations were not a problem.

\subsection{Self-learning Monte Carlo, EMUS, etc}
\begin{figure}[htp!]
	\centering
	\includegraphics[width=\columnwidth]{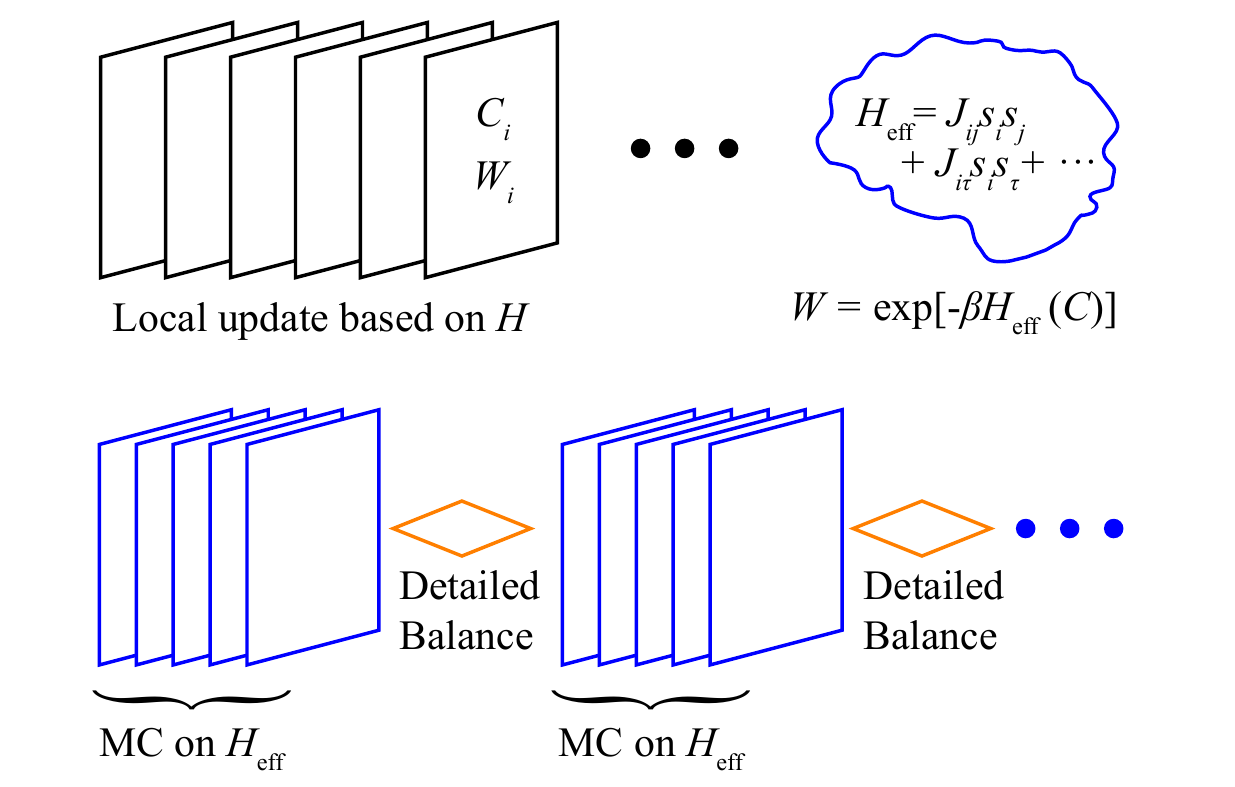}
	\caption{Schematic illustration of the learning process (top panel) and simulating
		process (bottom panel) in self-learning Monte Carlo. The fitting process happened inside the "brain" in the upper panel based on the configurations generated from the local update of the original Hamiltonian $H$, and once the effective Hamiltonian $H_{\text{eff}}$ is obtained, the cumulative updates of it are performed as shown in the lower panel. The detailed balance is guaranteed by the evaluation of the configurational weight according to the orginal Hamiltonian after cumulative update.}
	\label{fig:fig2}
\end{figure}

From the above discussion it is obvious that the evaluation of determinant inevitably renders the $O(\beta N^3)$ scales of operation, which is the computational bottleneck that governs whether DQMC could access larger system sizes and lower temperatures, and what make things worse is the critical slowing down inside the QCR. While physically critical regions are difficult as the fluctuations are the strongest here, but intelligent Monte Carlo update scheme can always help to greatly reduce the autocorrelation among the Monte Carlo configurations and effectively reduces the physical time spent. Successful examples include the Wolff and Swenden-Wang algorithms of the Ising model at classical critical points~\cite{Swendsen1987,Wolff1989} and loop-updates and stochastic series expansions in the quantum bosonic and spin systems~\cite{Evertz2003,Sandvik2003}. The successful cluster or non-local Monte Carlo updates all share the merit that the Monte Carlo move actually captures the correct modes in the low-energy landscape of the problem at hand, for example the Wolff cluster resembles the scale-invariance at the critical point and the size of the cluster varies in all length scales. It is even more so in the quantum Monte Carlo simulations in loop-updates or stochastic series expansion. Hence, for the spin-fermion model of our interest, it will be of crucial importance that a non-local update can be designed to respect the effective low-energy physics of the QCR and to overcome the autocorrelation and reduce the physical computational time. This goal, as will be explained here, is partially achieved by the self-learning Monte Carlo method and its subsequential developments such as elective momentum ultra-size Monte Carlo (EMUS) method.

The self-learning Monte Carlo update scheme (SLMC)~\cite{JWLiu2017SLMC,liu2016fermion,XYXu2017SLMC,Nagai2017,HTShen2018,
ChuangChen2018a} is designed to speed up the Monte Carlo simulations, both in classical~\cite{JWLiu2017SLMC,liu2016fermion,Bojesen2018}, quantum few-body~\cite{Nagai2017,HTShen2018} and quantum many-body~\cite{XYXu2017SLMC,ChuangChen2018a,ChuangChen2018b} systems. As shown in Fig.~\ref{fig:fig2}, in the context of strongly-correlated fermion systems, the SLMC is implemented in the following manner. One first performs the standard DQMC simulation on the model in Eq.~\eqref{eq:specificformula}, and then train an effective boson Hamiltonian that contains long-range two-body interactions both in spatial and temporal directions. The effective Hamiltonian serves as the proper low-energy description of the problem at hand with the fermion degree of freedom integrated out. We then use the effective Hamiltonian to guide the Monte Carlo simulations, i.e., we will perform many sweeps of the effective bosonic model (as the computational cost of updating the boson model is $O(\beta N)$, drastically lower than the update of fermion determinant which scales as $O(\beta N^3)$), and then evaluate the fermion determinant of the original model in Eq.~\eqref{eq:spin-fermion} such that the detailed balance
of the global update is satisfied. As shown in our previous works~\cite{JWLiu2017SLMC,XYXu2017SLMC,ZiHongLiuTri2018,ZiHongLiuEMUS2018,ZiHongLiuSqu2018,
ChuangChen2018a,ChuangChen2018b}, the SLMC can greatly reduce the autocorrelation time in the conventional DQMC simulation and make the larger systems and lower temperature accessible. 

\begin{figure}[htp!]
	\centering
	\includegraphics[width=\columnwidth]{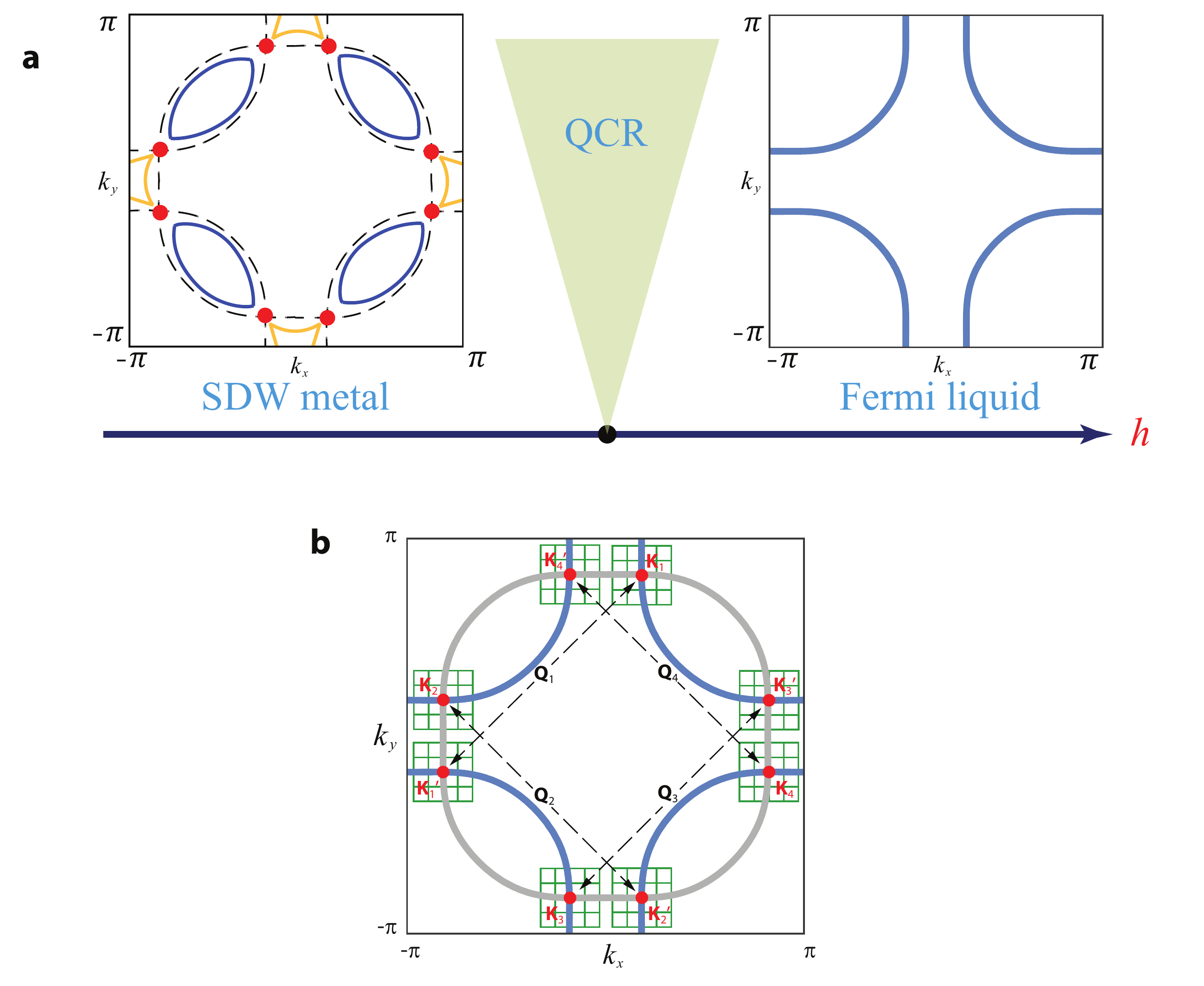}
	\caption{(a) Illustration of the fermionic quantum critical point with antiferromagnetic fluctuations on 2d square lattice. The lattice construction is the same as in Fig.~\ref{fig:fig1}(a), but the Ising spins (bosons) have antiferromagnetic interactions among them. Consequently, in the disordered phase of the bosons $(h>h_c)$, the fermions are in FL with the bare dispersion; in the ordered phase of bosons, the fermions form a spin-density-wave (SDW) metal with folded FS according to the $\mathbf{Q}=(\pi,\pi)$ fluctuations, the resulting FS is comprised of pockets and hot spots. In between the SDW metal and FL phase, lies the quantum critical region (QCR) where non-Fermi-liquid on the hot spots and the quantum critical scaling of the AFM-QCP are revealed with quantum Monte Carlo. (b) Momentum patches inside Brilliouin zone (BZ) of the model in (a). The blue lines are the Fermi surface (FS) of $H_{f}$ and $\mathbf{Q}_i=(\pm \pi, \pm \pi), \; i=1,2,3,4$ are the AFM wavevectors, and the four pairs of  $\{\mathbf{K}_i,\mathbf{K'}_i\}, \; i=1,2,3,4$ are the position of the hot spots (red dots), each pair is connected by a $\mathbf{Q}_i$ vector. The folded FS (gray lines), coming from translating the bare FS by momentum $\vec{Q}_i$. The green patches show the $\mathbf{k}$ mesh built around hot spots, number of momentum points inside each patch is denoted as $N_f$. Figure (b) is adapted from Ref.~\cite{ZiHongLiuSqu2018}.}
	\label{fig:fig3}
\end{figure}

As will be shown in the next section, SLMC worked very well for the spin-fermion models, and as we started to explore more related models, further numerical improvements, inspired by the theoretical understanding of the fermionic quantum critical points, were also made possible. Among them, elective momentum ultra-size quantum Monte Carlo method (EMUS) is of particular advantage in dealing with fermionic quantum critical point subject to antiferromagnetic or other finite momentum bosonic fluctuations. EMUS is developed by us in Ref.~\cite{ZiHongLiuEMUS2018}, it is inspired by the awareness that in a finite $\mathbf{Q}$ quantum critical point, the critical bosonic fluctuations mainly couple to fermions near the hot spots, which are the points on the FS connected by the finite $\mathbf{Q}$ bosonic fluctuations as shown in Fig.~\ref{fig:fig3} in the case of square lattice band structure subject to antiferromagnetic $\mathbf{Q}= (\pi,\pi)$ fluctuations. Thus, instead of including all the fermion degrees of freedom, we ignore fermions far away from the hot spots and focus only on momentum points near the hot spots in the simulation. This approximation will produce different results for non-universal quantities compared with the original model, such as critical field $h_c$ or critical temperature $T_c$, as one actually performed a hard UV cutoff to the problem at hand. However, for universal quantities at the IR, such as scaling exponents, which are independent of microscopic details and the high energy cutoff, EMUS has been shown to generate consistent values with those obtained from standard DQMC~\cite{ZiHongLiuEMUS2018,ZiHongLiuSqu2018}, both for the itinerant QCPs on triangle and square lattices.

In DQMC, configurations of the bosonic modes $\phi_i$ are stochastically sampled, with weights obtained through integrating out fermion modes. Conventionally, real-space single particle modes, $c_{ia}$, are chosen as the basis on which the calculation is performed. While in EMUS, we instead use the momentum space. To this end, we rewrite Eqs.~\eqref{eq:specificformula} in momentum space,
\begin{equation}
\label{eq:Hfk}
H_f=\sum_k[\epsilon(k)-\mu]c_{ka}^\dagger c_{ka};
\end{equation}
and
\begin{equation}
\label{eq:Hfbk}
H_{fb}=\lambda\sum_{kk^\prime}c_{ka}^\dagger M_{ab}c_{k^\prime b}
\phi_{k-k^\prime}.
\end{equation}
Here, $\phi_k$ denotes the $k$-component of the Fourier transform of the bosonic field $\phi_i$:
\begin{equation}
  \label{eq:phik}
  \phi_k = \frac1N\sum_i\phi_i e^{-i\bm k\cdot\bm r_i}.
\end{equation}
Rewriting the problem in the momentum space gives us a freedom of choosing arbitrary $k$ points in the summation in Eqs.~\eqref{eq:Hfk} and \eqref{eq:Hfbk}.
When to study the low-energy and long-wavelength physics, we choose IR fermion modes that are particularly relevant, and throw away other modes without worrying a proper UV completion of a lattice model.
Particularly, for studying AFM-QCP as shown in Fig.~\ref{fig:fig3} (a), only fermion modes near hot spots, where two patches of FSs are connected by the ordering wave vector $\mathbf{Q}_i$ [see Fig.~\ref{fig:fig3} (b)], are relevant to universalities. As usually found in the analytic calculation of such problems~\cite{Metlitski2010b,Abanov2004}, one keeps modes in patches around these hot spots, and neglect other modes in the BZ, as shown in the patches  in Fig.~\ref{fig:fig3} (b).
In this way, the number of fermion modes used in computing the effective weights is greatly reduced, from the total size (or total volume) $N$ to the patch size (or patch volume) $N_f$. Therefore, while retaining the same IR physics of what a lattice DQMC simulation of system size $N$ can achieve, EMUS has drastically lifted the computational burden.

In EMUS, as local couplings in real space become non-local in the momentum basis [Eq.~\eqref{eq:Hfbk}], making a simple local update in the standard DQMC costs $\beta N\cdot O(\beta N_f^3)$ computational complexity, one can no longer use such kind of local update. Fortunately, the cumulative update scheme in the SLMC developed~\cite{liu2016fermion,ZiHongLiuTri2018,ZiHongLiuEMUS2018,ZiHongLiuSqu2018,
ChuangChen2018a,ChuangChen2018b} come to help. Such a cumulative update is a global move of the Ising spins and gives rise to the  computational complexity at most $O(\beta N_f^3)$ for computing the fermion determinant. Since $N_f$ can be much smaller than $N$, a speedup of the order $(\frac{N}{N_f})^3 \sim 10^3$ of EQMC over DQMC, with $\frac{N}{N_f} \sim 10$, can be comfortably achieved.

\begin{figure*}[t]
	\includegraphics[width=\textwidth]{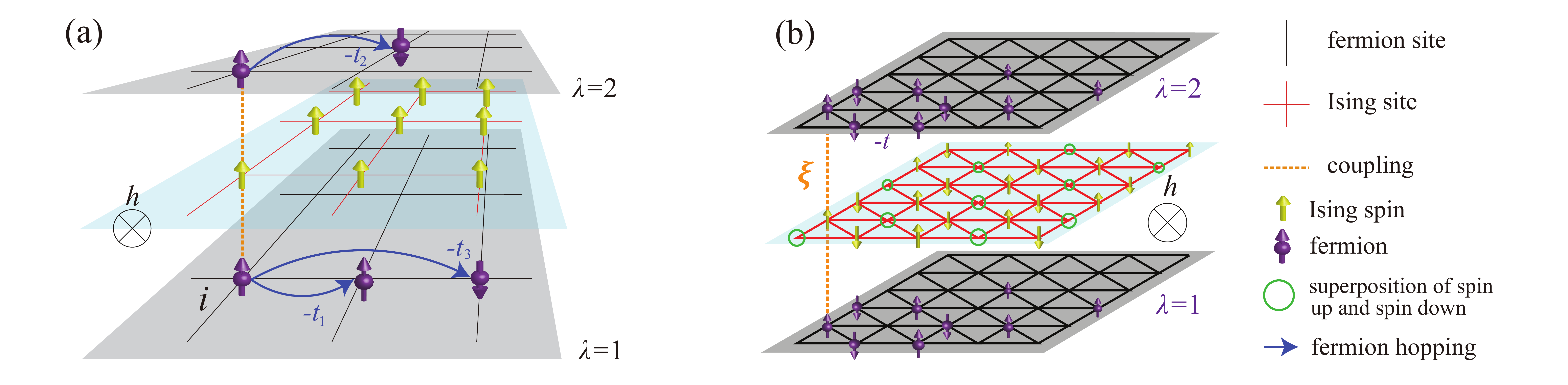}
	\caption{(a) Microscopic model of FM-QCP and AFM-QCP on square lattice. In FM-QCP, only nearest neighbor hopping $t$ is considered in each layer($t=t_1$), while in AFM-QCP, the nearest, next nearest and third nearest hopping ($t_1$, $t_2$, $t_3$) in each layer are all considered. (b) Microscopic model of AFM-QCP on triangular lattice. The discussions on the QMC results of these models are presented in Sec.~\ref{sec:Results}. Figure (a) is adapted from Refs.~\cite{Xu2017,ZiHongLiuSqu2018}, Figure (b) is adapted from Ref.~\cite{ZiHongLiuTri2018}. }
	\label{fig:fig4}
\end{figure*}

\section{Results}
\label{sec:Results}
In this section, we present several of our recent works on the fermionic quantum critical points, revealed with the quantum Monte Carlo methods discussed in the previous section. Most of these results can be found in the several recent publications, including the Ising ferromagnetic quantum critcal point (FM-QCP) studied in Ref~\cite{Xu2017}, the Ising antiferromagneitc quantum critical point (AFM-QCP) on the triangular lattice (with 3$\mathbf{Q}=\Gamma$) studied in Ref~\cite{ZiHongLiuTri2018} and the Ising antiferromagnetic quantum critical point on the square lattice (with $2\mathbf{Q}=\Gamma$) studied in Ref.~\cite{ZiHongLiuSqu2018}. From the point of view of a pedagogical narrative, here we took the content of each paper and restructured them in a coherent manner, such that the overall logical flow and the internal relation of the physical results beyond each individual papers manifest.
	

\subsection{Phase Diagrams}

\subsubsection{FM-QCP}
We start from the ferromagnetic quantum critical point (FM-QCP). In the general designer Hamiltonian illustrated in Eq.~\eqref{eq:specificformula}, we take $J<0$ to realize a FM-QCP and particularly, we are interested in the case where the Fermi surface away from perfect nesting which can be realized by tuning chemical potential. We draw the microscropic lattice model again in Fig.~\ref{fig:fig4} (a).

\begin{figure*}[t]
	\includegraphics[width=\textwidth]{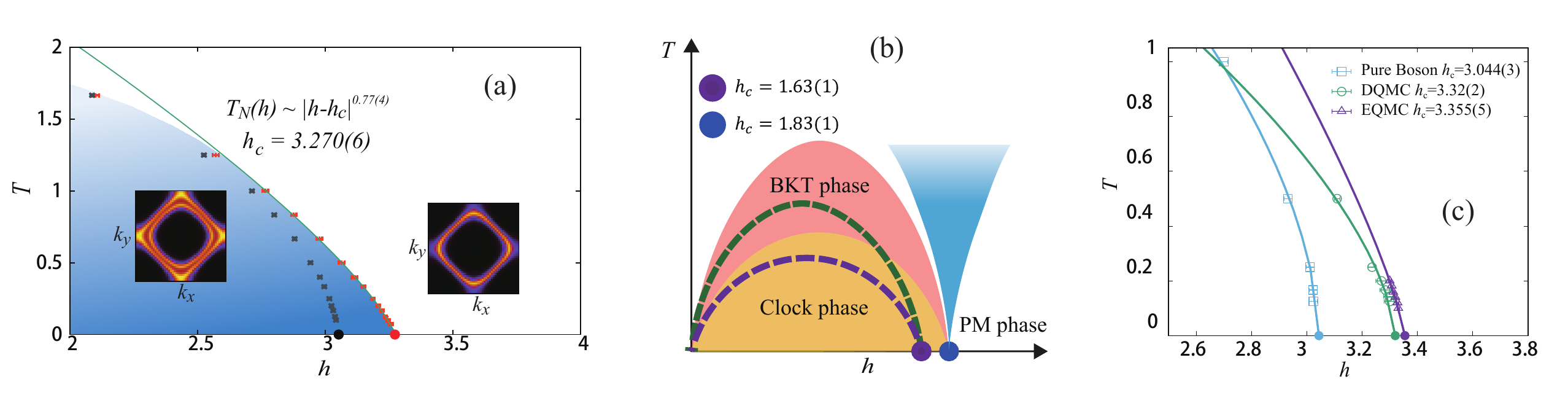}
	\caption{(a) Phase diagram of FM-QCP on square lattice. The splitting of the FS insides the FM ordered phase and its restructuring in the disordered phase are shown. The bare phase boundary of $H_{\text{boson}}$ (black dots and line) and the phase boundary of the coupled system $H_{\text{fermion}}+H_{\text{boson}}+H_{\text{coupling}}$ (red dots and line) are also shown. The $h_c=3.270(6)$ for the coupled system is larger than $h_c=3.04$ of the bare bosonic one. (b) Phase diagram of 3Q=$\Gamma$-AFM-QCP. The dashed lines are phase boundaries of $H_{\text{boson}}$ with a QCP (magena dot) at $h_c=1.63(1)$~\cite{YCWang2017}. The filled areas are phases of the coupled system with QCP (blue dot) shifted to a higher value $h_c=1.83(1)$. (c) Phase diagram of 2Q=$\Gamma$-AFM-QCP. The light blue line is AFM-PM phase boundary of $H_{\text{boson}}$ with a QCP (light blue dot) at $h_c=3.044(3)$. The green line is phase boundary of couple system obtained with DQMC (QCP at $h_c=3.32(2)$, green point), while violet line obtained with EMUS (QCP at $h_c=3.355(5)$, violet point). Figure (a) is adapted from Ref.~\cite{Xu2017}, (b) from Ref.~\cite{ZiHongLiuTri2018}, and (c) from Ref.~\cite{ZiHongLiuSqu2018}.}
	\label{fig:fig5}
\end{figure*}

In the simulation, we set $t=1$, $J=-1$ and  chemical potential $\mu=-0.5$, resulting in a fermion density $\langle n_{i\lambda}\rangle \approx 0.8$, with FS away from perfect nesting. The coupling strength between the fermion and bosons is set to $\xi=1$. By tuning the transverse field $h$, we realized an itinerant PM-FM transition as shown in the phase diagram of Fig.~\ref{fig:fig5}(a). In the temperature-transverse filed ($T-h$) phase diagram, the black dots are the phase boundary of bare transverse Ising model without coupling to fermions ($\xi=0$ case) where  the finite temperature paramagnetic (PM) to ferromagnetic (FM) phase transition belongs to 2D Ising universality class, and the zero temperature PM-FM transition (large black dot) belongs to 3D Ising universality class. When turning on the couplings between the Ising spins and fermions ($\xi=1$ here), the phase boundary shift a little bit to larger $h$ direction (red points), and the finite temperature PM-FM phase transition is still 2D Ising as at finite temperature the Fermion can be trivially integrated out without influencing the criticality, while at zero temperature (large red point), there exist no controllable analytical tools to take care of critical fluctuations of these strongly coupled fermions and bosons, and the criticality here is believed to be different from 3D Ising universality and in fact different from any known ones. Our QMC numerical results reveal the properties of such itinerant 3D FM-QCP and indeed discovered the scaling behavior near the QCP with new critical exponents significantly deviate from 3D Ising with critical FS of non-FL and large anomalous dimensions, as will be discussed in Secs.~\ref{sec:nFL} and \ref{sec:QCscaling}. 

\subsubsection{AFM-QCP with $3\mathbf{Q}=\Gamma$}
We choose a triangular lattice to realized a AFM-QCP with $3\mathbf{Q}=\Gamma$-by setting $J>0$ in the general designer Hamiltonian in Eq.~\eqref{eq:specificformula}, as shown in Fig.~\ref{fig:fig4} (b). The notation of $3\mathbf{Q}=\Gamma$ means that three times length of the antiferromagnetic fluctuations $\mathbf{Q}=(\frac{2\pi}{3},\frac{2\pi}{\sqrt{3}})$ will bring the wavevector back to the $\Gamma$ point in the extended BZ scheme. Once we apply the same mode on square lattice, it is obvious that twice length of the $\mathbf{Q}=(\pi,\pi)$ wavevector will bring the fluctuations back to $\Gamma$ in the extended zone, therefore dubbed AFM-QCP with $2\mathbf{Q}=\Gamma$. As will become clear later in Sec.~\ref{sec:QCscaling}, these two types of AFM-QCPs actually have different scaling behaviors, consistently revealed both from QMC simulation and theoretical argument.

We set $t=1$, $J=1$, $\mu=-0.5$ (electron density $\langle n_{i,\lambda} \rangle \sim 0.8$ and explore the $T$-$h$ phase diagram as shown in Fig.~\ref{fig:fig5}(b). When the Ising spin and fermion is not coupled ($\xi=0$ case), the phase diagram is already well studied~\cite{Moessner2001,Isakov2003,YCWang2017}, and the phase boundary is denoted as dashed lines Fig.~\ref{fig:fig5}(b). At low temperature and small $h$, a clock phase is formed, which is characterized by a complex order parameter $me^{i\theta}=m_{1}+m_{2}e^{i4\pi/3}+m_{3}e^{-i4\pi/3}$ where $m_{\alpha}$, with $\alpha=1,2,3$, equals to $\frac{1}{3N}\sum^{N/3}_{i=1} s^{z}_{i,\alpha}$, representing magnetization of the three sublattices. 
In clock phase, the mentioned order parameter has a finite momentum $\mathbf{Q}=(\frac{2\pi}{3},\frac{2\pi}{\sqrt{3}})$. At $T=0$, increasing $h$ destroys clock phase through a continuous quantum phase transition at $h_c=1.63(1)$ and a U(1) symmetry is emerged at QCP, making the quantum phase transition guided by the (2+1)D XY universality class~\cite{Isakov2003,YCWang2017}. When increasing temperature, the clock phase finally melts and thermal melting goes through an intermediate BKT phase. When the Ising spin fermion is coupled ($\xi=1$ case), the phase boundary shifted to a larger ordered region and the QCP is now at $h_c=1.83(1)$ as shown in Fig.~\ref{fig:fig5}(b). As the Ising spin clock phase generates a SDW order (with wavevector $\mathbf{Q}$) in the fermionic sector, the fermion band is folded with six FS pockets and six pairs of hot spots connected by the $\mathbf{Q}$ wavevector (the detailed FS structure can be seen in Fig.1 of Ref.~\cite{ZiHongLiuTri2018}). In contrast to FM QCP case with a $\mathbf{Q}=0$ where the entire FS is critical, here with a finite momentum $\mathbf{Q}$ QCP, only the hot spots are critical, and we saw non-Fermi liquid behavior at hot spots, as will be discussed in Sec.~\ref{sec:nFL} and Sec.~\ref{sec:QCscaling}. We have performed complementary DQMC and EMUS simulations on the model, as shown in Ref.~\cite{ZiHongLiuTri2018} and Ref.~\cite{ZiHongLiuEMUS2018} respectively.

\subsubsection{AFM-QCP with $2\mathbf{Q}=\Gamma$}
\label{sec:AFMFS}

Antiferromagnetic quantum critical point can be realized in many forms. Here we implement it in the form of square lattice with $\mathbf{Q}=(\pi,\pi)$ fluctuations. We realized a $2\mathbf{Q}=\Gamma$ AFM-QCP by setting $J>0$ in the general designer Hamiltonian in Eq.~\eqref{eq:specificformula} and the microscopic lattice model is depicted in Fig.~\ref{fig:fig4}(a).

To tune the FS, in particularly to make it represent that of the high temperature superconductor cuprates, we design fermion hopping part as
\begin{align}
H_{\text{fermion}}  = & -t_1\sum_{\left \langle ij \right \rangle, \lambda ,\sigma }c_{i,\lambda, \sigma}^\dagger c_{j,\lambda ,\sigma}-t_2\sum_{\left \langle \left \langle ij \right \rangle \right \rangle, \lambda ,\sigma }c_{i,\lambda ,\sigma}^\dagger c_{j,\lambda ,\sigma}\nonumber \\
& -t_3\sum_{\left \langle \left \langle \left \langle ij \right \rangle \right \rangle \right \rangle ,\lambda ,\sigma }c_{i,\lambda ,\sigma}^\dagger c_{j,\lambda, \sigma}+h.c.-\mu\sum_{i,\lambda ,\sigma}n_{i,\lambda ,\sigma} ,
\end{align}
and choose $t_1=1.0$, $t_2=-0.32$, $t_3=0.128$, $\mu=-1.11856$ (electron density $\langle n_{i,\lambda}\rangle \sim 0.8$), according to Ref.~\cite{Chowdhury2014}.

\begin{figure}[htp!]
\centering
\includegraphics[width=\columnwidth]{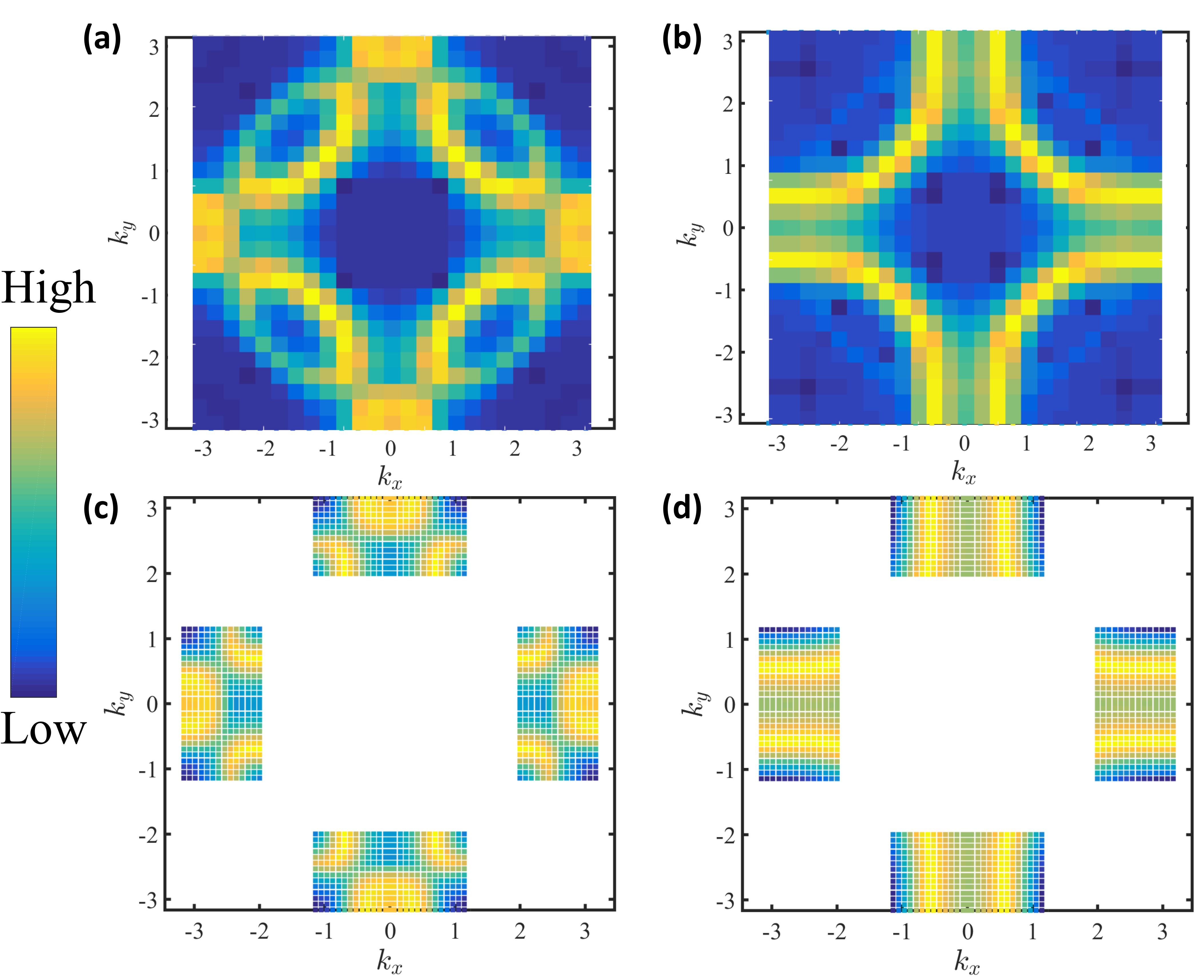}
\caption{FS obtained from DQMC (panels (a) and (b)) and EMUS (panels (c) and (d)). The FS is obtained from fermion spectrum function at zero energy $A(\mathbf{k},\omega=0)$ utilizing the standard approximation $G(\mathbf{k},\beta/2) \sim A(\mathbf{k},\omega=0)$. (a) and (c), FS in the AFM ordered phase ($h<h_c$), where Fermi pockets are formed from zone folding. DQMC and EMUS results are consistent with each other, while EMUS (with system size $L=60$) gives much higher resolution in comparison with DQMC ($L=28$). (b) and (d), similar comparison at the QCP ($h=h_c$). Figures adapted from Ref.~\cite{ZiHongLiuSqu2018}.}
\label{fig:fig6}
\end{figure}

We set $J=1$ and implement both DQMC and EMUS here to explore the $T-h$ phase diagram as shown in Fig.~\ref{fig:fig5}(c). The AFM-QCP with $2\mathbf{Q}=\Gamma$ is realized by turning on Ising spin fermion coupling (we set $\xi=1$ here). The standard DQMC with SLMC update scheme greatly reduce the autocorrelation time and make the larger systems and lower temperature accessible. We use it provides unbiased results as a benchmark. On the other hand, EMUS concentrates on fermions with momentum near the hot spots, therefore it enjoys  a  speed up of the order $( \frac{N}{N_f} )^3$ and makes even larger system size accessible. Fig.~\ref{fig:fig6} shows FS obtained by $G(\mathbf{k},\beta/2)\sim A(\mathbf{k},\omega=0)$ with both DQMC and EMUS. Higher momentum resolution near hot spots is obtained in EMUS, and it plays a vital role in unveiling critical behavior of this QCP. 

As will be discussed in detail in Sec.~\ref{sec:nFL} and Sec.~\ref{sec:QCscaling}, although the AFM-QCP with $3\mathbf{Q}=\Gamma$ and $2\mathbf{Q}=\Gamma$ share similarities, such as the formation of FS pockets and hot spots. They give rise to very different quantum criticalities. It turns out the length of the antiferromagnetic wavevector actually matters in terms of the effects of the critical fluctuations. In the triangular lattice case, the QCP is found to be consistent with the HMM prediction, within our current energy and momentum resolution. However, in the square lattice case, quantum criticality with large anomalous dimension is discovered with scrutiny. The discovery of anomalous dimension is of great significance in the theoretical understanding of the fermionic quantum critical points. As many field theoretical developments beyond the HMM, such as those in Refs.~\cite{Abanov2003,Abanov2004,Metlitski2010b,Metzner2012,Schlief2017,SSLee2018}, the anomalous dimension at the AFM-QCP (actually also the FM-QCP but it is even more difficult to handle analytically) have been suggested to have various forms, such as a function of the number of hot spots $N_{\text{h.s.}}$, but the validity of such renormalization group (RG) calculations are not known a priori, since it is still the fundamental question that a proper small parameter for the RG flow for fermionic quantum critical point is in absence. Therefore, our quantum Monte Carlo results actually provide the first set of unbiased attempts to provide concrete results for the analytical approaches to compare with, and from such comparison, further adjustments from both theoretical and numerical sides, can be made to reach the eventually solution of the problem. In this regard, our results shown in the review, in particular in Sec.~\ref{sec:QCscaling}, will be of great importance for the further developments of the community.

\subsection{Non-Fermi liquid behavior}
\label{sec:nFL}

\begin{figure}[htp!]
\includegraphics[width=\columnwidth]{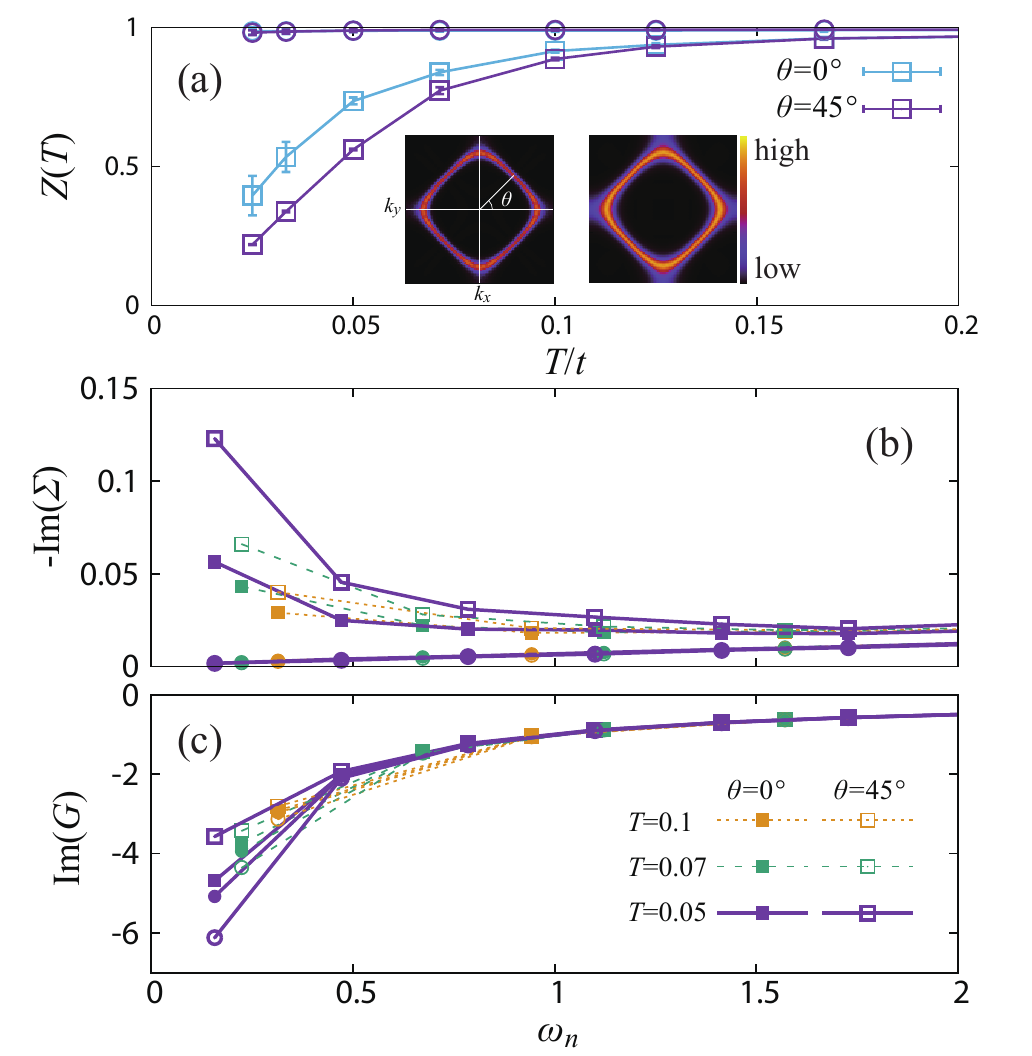}
\caption{(a) $Z_{\mathbf{k}_{F}}(T)$ at FM-QCP ($h_c=3.27$, squares) and in PM phase ($h=3.60$, circles). The left inset is the $G(\mathbf{k},\frac{\beta}{2})$ at FS for $T=0.05$ while the right inset is for $T=0.1$. Although there is anisotropy in $Z_{\mathbf{k}_{F}}$ at different parts of the FS, the quasi-particle weight in $k_x$ and $k_x=k_y$ directions both approach zero at FM-QCP, indicating a nFL behavior for the entire FS.  The data in the PM phase shows the quasi-particle weight approaching a constant (very close to 1), indicating the system is a FL. (b) $-\text{Im}(\Sigma(\mathbf{k}_{F},\omega_n))$ at FM-QCP ($h=h_c$, square symbol), it increases as $\omega_n \to 0$ -- signifying the system at QCP loses their quasi-particle weight with a power law -- a nFL behavior, while in PM phase ($h=3.60$, circle symbol) the imaginary part of self energy approaches zero linearly as $\omega_n \to 0$ -- a FL behavior.
(c) Imaginary part of the single-fermion Green's function at the FM-QCP  ($h=3.27$, square symbol) and in PM phase  ($h=3.60$, circle symbol). No signature of gap formation is observed. Figures adapted from Ref.~\cite{Xu2017}.}
\label{fig:fig7}
\end{figure}

Now let's focus on how the fermionic degrees of freedom react at the fermion QCPs. In particular, we would like to find out whether nFL behavior with the low-temperature fermionic quasi-particle weight vanishes at the Fermi surface or hot spots~\cite{Oganesyan2001,Senthil2008,Lee2009,Metlitski2010a,Metlitski2010b,Dalidovich2013,Schlief2016,Lee2017} due to strong damping of the fermions induced by critical bosonic fluctuations, can be seen from our simulations.
Following~\cite{Chen2012}, the first Matsubara frequency ($\omega_0=\pi T$) of self-energy gives a good estimation of quasi-particle weight 
\begin{equation}
Z_{\mathbf{k}_{F}} \approx \frac{1}{1-\frac{\text{Im}\Sigma(\mathbf{k}_{F},i\omega_{0})}{\omega_{0}}},
\label{eq:Z}
\end{equation}
as shown in Fig.~\ref{fig:fig7}(a) for $h=h_c$ (squares) and $h>h_c$ (circles) for the FM-QCP case. The behavior of quasi-particle weight with temperature in the PM phase ($h>h_c$) and at QCP ($h=h_c$) is totally different, that it remains close to unity at low temperature in the PM phase, indicating well-defined quasi-particles, while it is strongly suppressed and extrapolates to zero as $T\rightarrow 0$ at QCP, which is a key signature of a non-Fermi liquid. The different quasi-particle behaviors in the PM phase and at QCP is also revealed in the fermions self-energy and Green's function data as shown in Fig.~\ref{fig:fig7}(b) and (c). The increasing of both $-\text{Im}(\Sigma)$ and $-\text{Im}(G)$ with decreasing $\omega_n$ at QCP indicates strong damping of the fermions at low frequencies but without any signature of gap opening. Such enhancement of the self-energy clearly suggests the strong correlation at the QCP, which means either this is a new fermionic QCP without previously knowledge or that when the temperature/energy scales go lower than what have accessed by now the system might develop a first order transition~\cite{Belitz2005}. But either the case is interesting and have not been addressed with any unbiased approaches before.

In the FM-QCP, since the bosonic fluctuations are at $\mathbf{Q}=0$, the entire FS become critical and as shown in Fig.~\ref{fig:fig7} the quasiparticles indeed lose coherence at the FM-QCP. The AFM-QCPs, in this regard, are different and our results show that it is only at the hot spots points, i.e., the momentum points connected by the antiferromagnetic wavevectors, the nFL behavior can be observed at the AFM-QCP, as shown in Fig.2 in Ref.~\cite{ZiHongLiuTri2018} for triangular lattice case and in Fig.4 in Ref.~\cite{ZiHongLiuSqu2018} in the square lattice case. But at the momenta away from the hot spots, the FL behavior of the quasiparticle keep intact as their quasiparticle fraction measured in Eq.~\eqref{eq:Z} remains close to 1. From these results, one could also tell that actually the FM-QCP is harder to handle than the AFM-QCP in that at the FM-QCP all the fermionic modes on the FS are strongly coupled to the critical bosonic fluctuations and it is therefore difficult to develop controlled RG calculation scheme. At the AFM-QCP, only finite number of hot spots fermionic modes are critically coupled with finite $\mathbf{Q}$ bosonic fluctuations (in the case in Sec.~\ref{sec:AFMFS}, $N_{\text{h.s.}}=16$).
As will be seen in Sec.~\ref{sec:QCscaling}, a finite $\mathbf{Q}$ QCP is theoretically better controlled in comparison to QCPs with $\mathbf{Q}=0$, and recent numerical results show good agreement with theory predictions at the qualitative level. However, at the quantitative level, numerical results still differ from theory predictions by a numerical factor, which requires further investigations and efforts from both numerical and analytical sides.

\subsection{Quantum critical scaling analysis}
\label{sec:QCscaling}

After showing the fermionic degrees of freedom at the quantum critical points, we now move on to the most difficult and yet the most direct observable that reveal the nature of the QCP -- the dynamic magnetic susceptibilities. In fact, all the previous analytical works, are in one way or the other, trying to find controlled scheme to integrate out the fermionic degrees freedom at the quantum critical point and arrive at a bosonic susceptibility that captures the unique divergence at the QCP. Therefore, our results in this section provide solid examples that analytical approaches can be used to test with and might inspire further analytical developments.
 
\subsubsection{FM-QCP}
We again start with the FM-QCP. Our results in Ref.~\cite{Xu2017} show that the critical behavior of itinerant FM-QCP is strongly deviates from 3D Ising universality class. The first evidence comes from the PM-FM transition temperature, as shown in Fig.~\ref{fig:fig5}(a), the phase boundary of the FM onset temperature scales as $T_N(h)\sim |h-h_c|^c$. We have $c=0.77(4)$ for the FM-QCP case while that for 3D Ising is $c=\nu z \approx 0.63$, also note that due to itinerant nature of the QCP, the exponent $c$ is no longer expected to obey the relation $c=\nu z$~\cite{Loehneysen2007,Belitz2005}.

A more rigorous analysis of scaling behavior is based on the scaling analysis of the data inside the QCR, in which the bosonic critical modes become strongly renormalized due to the coupling to gapless fermions. Our QMC results indicate that the scaling behavior deviates from the HMM predictions significantly and we propose a modified Hertz-Millis scaling formula which fits our QMC data for the Ising spin susceptibility at all the momenta and frequencies simulated near QCP in the long wavelength and low frequencies limit.

As discussed in Sec.~\ref{sec:HMM}, even at the level of random phase approximation (RPA), non-trivial quantum dynamics has already entered the bosonic susceptibility~\cite{Hertz1976,Millis1993,Moriya1985}. We, however, will measure the quantity unbiasedly in the quantum Monte Carlo simulations, the Ising spin susceptibility is given by,
\begin{equation}
\chi(h,T,\vec{q},\omega_n) = \frac{1}{L^2}\int_{0}^{\beta} d\tau \sum_{ij} e^{i\omega_n\tau - i\mathbf{qr}_{ij}} \langle s_i^z(\tau) s_j^z(0) \rangle.
\end{equation} 
At the RPA level, $\chi(h,T,\vec{q},\omega_n)$ has the following hypothesized form near the QCP,
\begin{align}
\chi(h,T,\vec{q},\omega_n)=\frac{1}{c_t T^2+c_h \left | h-h_{c} \right |+ c_q q^2+ c_\omega \omega^2+ \Delta(\vec{q},\omega_n)}.
\label{eq:HZ_chi}
\end{align}
where $c_t$, $c_h$, $c_q$, $c_\omega$ are constants. Here, the $c_q q^2+ c_\omega \omega^2$ comes from the bare action of the Ising
degrees of freedom, and the $\Delta(\vec{q},\omega_n)$ term is the contribution of the fermionic fluctuations, and has following form
\begin{align}
\Delta(\vec{q},\omega_n)=c_{\mathrm{HM}} \frac{|\omega_n|}{\sqrt{\omega_n^2+(v_f q)^2}}
\label{eq:Delta_RPA}
\end{align}
in the isotropic and low momentum/energy limit. The $c_{\mathrm{HM}}$ is a constant and $v_f$ is the Fermi velocity. The form in Eq.~\eqref{eq:Delta_RPA} has the following meaning that due to the coupling between critical bosons with fermions, the limits of $q=0$, $\omega \rightarrow 0$ and $\omega=0$, $q\rightarrow 0$ no longer commute with each other, which also points out the $z=1$ Lorentz symmetry present at the bare 3D Ising boson critical point no longer hold in the coupled case, that the FM-QCP is indeed a new quantum criticality. Such statement will be verified with our numerical data.

\begin{figure}
\includegraphics[width=\columnwidth]{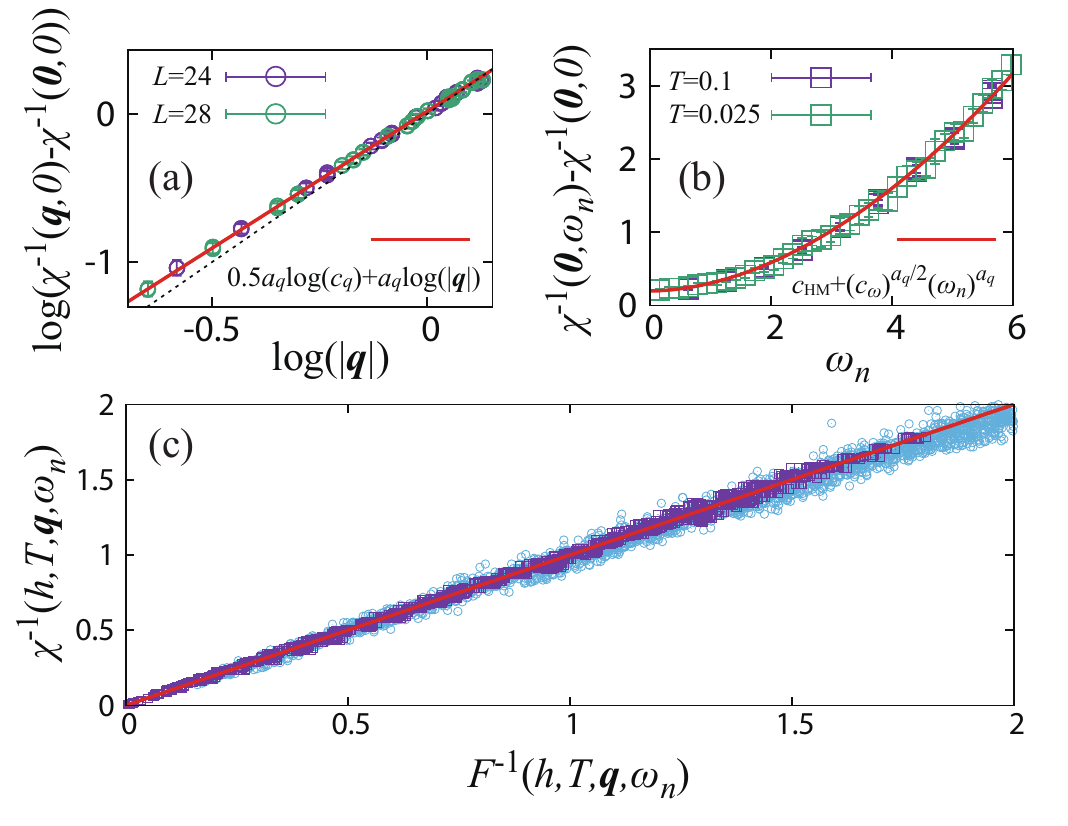}
\caption{
(a) Inverse Ising spin susceptibility at $\omega_n=0$ as a function of $|\vec{q}|$ (data points with $L=24$, $28$ and $T=0.125$). The red line shows the fitting with $\chi^{-1}=c_q q^{a_q}$ and we get $a_q=1.85(3)$. The black dashed line shows the slope $a_q=2$.
(b) Inverse Ising spin susceptibility at $q=0$ as a function of $\omega_n$ (data points with $L=20$, $24$ and $28$ for  $T=0.1$, with $L=20$ for $T=0.025$). The red curve shows the fitting with $\chi^{-1}=c_{\mathrm{HM}}+c_{\omega} \omega_n^{a_\omega}$.
(c) Data collapse for Ising susceptibility against the functional Eq.~\eqref{eq:HZ_modified_chi}, where
$F^{-1}=c_t T^{a_t}+c_h \left | h-h_{c} \right |^{\gamma}+ \left(c_q q^2+ c_\omega \omega^2\right)^{a_q/2}
+ \Delta(\vec{q},\omega_n)$.
The dark violet square points (3946 in total) are comprised of data with zero frequency while the light blue circle points are for data with frequency dependence. Figures adapted from Ref.~\cite{Xu2017}.}
\label{fig:fig8}
\end{figure}

Our QMC results share some characteristics with the above form, such as the singular behavior that $\lim_{q\to 0}\lim_{\omega_n\to 0} \chi^{-1}(\vec{q},\omega_n)$ differs from
$\lim_{\omega_n\to 0} \lim_{q\to 0}\chi^{-1}(\vec{q},\omega_n)$ by a constant $c_{\mathrm{HM}}=0.20(4)$, but importantly, we found that the RPA form shall be adjusted to a form with finite anomalous scaling dimensions,
\begin{align}
&\chi(h,T,\vec{q},\omega_n) \nonumber
\\
&=\frac{1}{c_t T^{a_t}+c_h \left | h-h_{c} \right |^{\gamma}+ \left(c_q q^2+ c_\omega \omega^2\right)^{a_q/2}
+ \Delta(\vec{q},\omega_n)},
\label{eq:HZ_modified_chi}
\end{align}
with $a_q = 1.85(3)$ ($\eta=0.15(3)$ from $a_q=2-\eta$) . This is different both from the exponent for an Ising transition in $(2+1)$ dimensions, $a_q=1.96$ ($a_q=2-\eta$ and $\eta=0.04$ for 3D Ising), and from the HMM RPA value of $a_q=2$. The presence of an anomalous exponent is another  key finding of the study in Ref.~\cite{Xu2017}. The singular behavior that in the limits of $q = 0$, $\omega \rightarrow 0$ and $\omega=0$, $q\rightarrow 0$ and that they do not commute are also found in a closely related quantum Monte Carlo investigation of nematic QCP~\cite{Schattner2016}, but due to the limitation of system size and temperature, the errorbar in the susceptibility in Ref.~\cite{Schattner2016} is too large (at the level of 0.3) to make any concrete statement on the absence and existence of the anomalous dimension we find here.

We fit all the constants and exponents in Eq.~\eqref{eq:HZ_modified_chi} based on our QMC data. For example, when setting $q=0$ and $\omega_n=0$ and follows the temperature and magnetic field dependence, we can obtain $c_t=0.13(1)$, $a_t=1.48(4)$, $c_h=0.7(1)$ and
$\gamma=1.18(4)$, these results can be seen in Fig.8 and 9 in Ref.~\cite{Xu2017}. Then one can set $\omega=0$, $h=h_c$ and at a very low temperature, to monitor the momentum dependence of $\chi(q)$ in Eq.~\eqref{eq:HZ_modified_chi}, the results are shown in Figs.~\ref{fig:fig8} (a). From two different system sizes, we can obtain $a_q=1.85(3)$ and $c_q=1.00(2)$. The finite $a_q$ suggests a finite anomalous dimension $\eta=0.15(3)$. In the same vein, one can set $q=0$, $h=h_c$ and at a very low temperature, to monitor the frequency dependence of the $\chi(\omega)$ in Eq.~\eqref{eq:HZ_modified_chi}, the results are shown in Fig.~\ref{fig:fig8} (b), from the fit, we obtain $c_{\omega}=0.10(2)$ and $c_{\text{HM}}=0.20(4)$. The finite $c_{\text{HM}}$ indicate the two limits $q = 0$, $\omega \rightarrow 0$ and $\omega=0$, $q\rightarrow 0$ do not commute, at least upto the system sizes and temperature accessed within our QMC simulation.

One way to show that the above fitting is robust is to use the obtained exponents and coefficients to collapse all $\chi$ data
for all $q$, $\omega_n$, $h$ and $T$ simulated according to Eq.~\eqref{eq:HZ_modified_chi}. Such a collapse is shown in Fig.~\ref{fig:fig8} (c), where all the data collapse onto the diagonal line, especially for small $q$, $\omega_n$, low temperature $T$ and $h\sim h_c$. Such a collapse strongly suggests that the FM-QCP revealed with our finite size QMC simulations is a novel quantum critical point and the results obtained will be useful for future analytical works to test with.

One more thing we want to remark here is about the exact form of  $\Delta(\vec{q},\omega_n)$, we found that as long as $\lim_{\omega_n\to 0} \Delta(\mathbf{q},\omega_n)=0$
and $\lim_{q\to 0}\Delta(\mathbf{q},\omega_n)=c_{\mathrm{HM}}=0.20(4)$ is kept, the impact of specific functional form of 
$\Delta(\vec{q},\omega_n)$ is little within numerical error bars of the QMC data, and
this uncertainty in $\Delta(\vec{q},\omega_n)$ implies that the exact form of $\Delta(\vec{q},\omega_n)$ can be much more complicated than the simple RPA approximation and the current system sizes and temperature are still not sufficient to distinguish their differences (e.g., the precise value of the dynamical critical exponent $z$). To further explore the nature of the FM-QCP, future simulations with larger system sizes and lower temperatures are still necessary and will be performed with even better Monte Carlo techniques, such as updated version of self-learning Monte Carlo. Fortunately, this does not affect our analysis at the static limit ($\omega_n=0$), and all the conclusions in this limit, including the anomalous dimensions $\eta=2-a_q=0.15(3)$, stand without any dependence on the details of $\Delta(\vec{q},\omega_n)$.

\begin{figure}[htp!]
\centering
\includegraphics[width=\columnwidth]{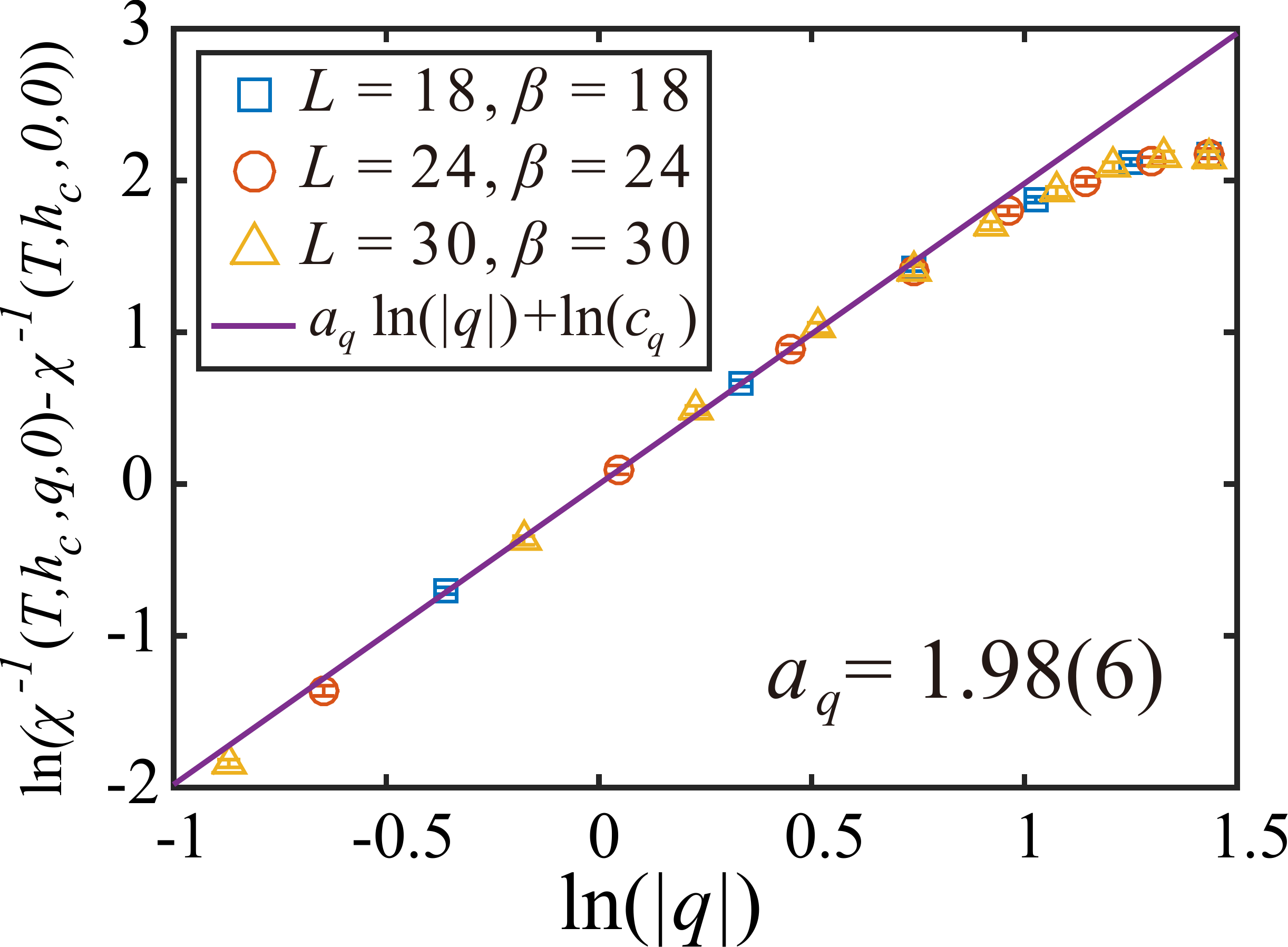}
\caption{At $h=h_c$, $\omega_n=0$, the plot of $\ln(\chi^{-1}(h_c,T,|\mathbf{q}|,0)-\chi^{-1}(h_c,T,0,0))$ as function of $\ln(|\mathbf{q}|)$ to obtain the power-law behavior $c_q|\mathbf{q}|^{a_q}|$ in the momentum dependence of bosonic susceptibility. Figure adapted from Ref.~\cite{ZiHongLiuTri2018}.}
\label{fig:fig9}
\end{figure}

\subsubsection{AFM-QCP with $3\mathbf{Q}=\Gamma$}
\label{sec:AFMQCP3Q}
For AFM-QCP with finite $\mathbf{Q}$, there is no singularity behavior under different orders of $\omega \rightarrow 0$ and $q \rightarrow 0$ limits, such equality applies for both $3\mathbf{Q}=\Gamma$ and $2\mathbf{Q}=\Gamma$ cases. For the former, we found the spin susceptibility can be well captured by the form
\begin{align}
&\chi(T,h,\mathbf{q},\omega_n)= \nonumber\\
& \frac{1}{(c_{t}T+c_{t}' T^{2})+c_{h}|h-h_c|^{\gamma}+c_q |\mathbf{q}|^2 + (c_{\omega}\omega+c'_{\omega}\omega^{2})}.
\label{eq:susceptibility}
\end{align}
At low temperature and frequency, the linear $T$ and $\omega$ terms dominant, this is in consistent with HMM prediction for antiferromagnetic cases. At higher temperature and frequency, the quadratic $T^2$ and $\omega^2$ terms dominant, in consistent with bare boson (2+1)D XY universality class for the antiferromagnetic Ising transverse field model on triangular lattice~\cite{Moessner2001,YCWang2017,Isakov2003}. Such crossover behavior from low to high energy scale is well established in Eq.~\eqref{eq:susceptibility} and in our QMC data. One thing we want to highlight here is we do not find anomalous dimension (see Fig.~\ref{fig:fig9}) in the $3\mathbf{Q}=\Gamma$ AFM-QCP case. As discussed in Sec.~\ref{sec:HMM} and will be discussed in Fig.~\ref{fig:fig11} in Sec.~\ref{sec:AFMQCP2Q}, this is consistent with the understanding that an important scattering channel is missed in this case, whose effect will be seen in the $2\mathbf{Q}=\Gamma$ AFM-QCP case. Our Monte Carlo simulation results in this and next sections are the first set of unbiased numerical data to reveal such interesting and subtle difference among the AFM-QCPs.

\begin{figure}[htp]
\centering
\includegraphics[width=\columnwidth]{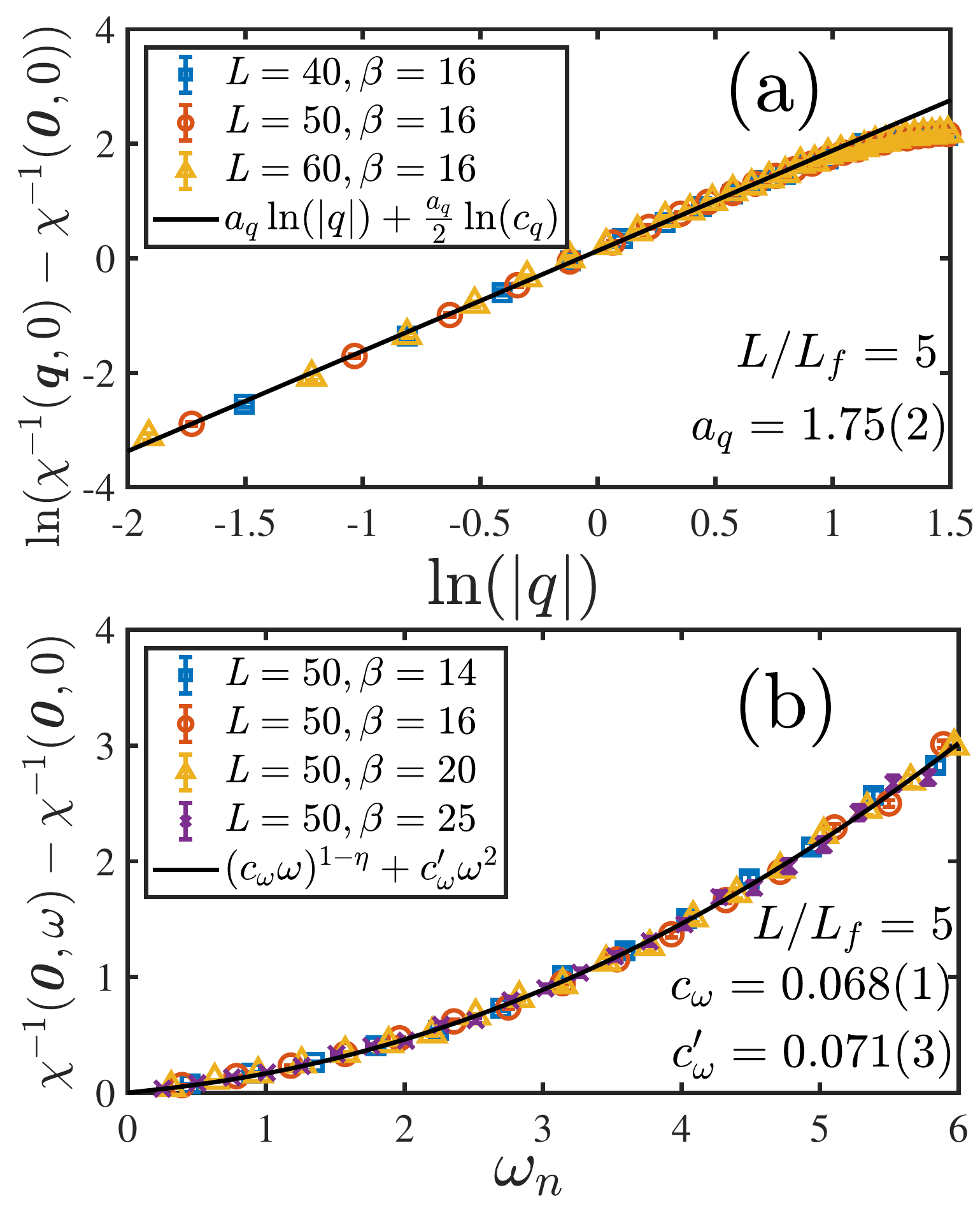}
\caption{(a) Momentum dependence of the bosonic susceptibilities $\chi(T=0,h=h_c,\mathbf{q},\omega=0)$ at the AFM-QCP. The system sizes are $L=40,50$ and $60$. The fitting line according to the form in Eq.~\eqref{eq:AFMQCPSQUARE} reveals that there is anomalous dimension in $\chi^{-1}(\mathbf{q})\sim |\mathbf{q}|^{2(1-\eta)}$ with $\eta=0.125$. (b) Frequency dependence of the bosonic susceptibilies $\chi(T=0,h=h_c,\mathbf{q}=0,\omega)$ at the AFM-QCP. The system size is $L=50$ and the temperature is as low as $\beta=25$ ($L_{\tau}=500$). The fitting line according to the form in Eq.~\eqref{eq:AFMQCPSQUARE} reveals that there is anomalous dimension in $\chi^{-1}(\omega) \sim \omega^{(1-\eta)}$ at small $\omega$ and crossover to $\chi^{-1}(\omega) \sim \omega^2$ at high $\omega$. Figures adapted from Ref.~\cite{ZiHongLiuSqu2018}.}
\label{fig:fig10}
\end{figure}

\subsubsection{AFM-QCP with $2\mathbf{Q}=\Gamma$}
\label{sec:AFMQCP2Q}
The scaling analysis of AFM-QCP with $2\mathbf{Q}=\Gamma$ is presented in this section. Comparing with the AFM-QCP with $3\mathbf{Q}=\Gamma$ in Sec.~\ref{sec:AFMQCP3Q}, AFM-QCP with $2\mathbf{Q}=\Gamma$ shows interesting features deviated further away from HMM. We found the magnetic susceptibility with the following form with an anomalous dimension can guide our fitting of QMC data very well.
\begin{align}
& \chi(T,h,\mathbf{q},\omega_n) \nonumber \\
& =\frac{1}{c_{t}T^{a_t}+c_{h}|h-h_c|^{\gamma}+(c_q |\mathbf{q}|^{2} + c_{\omega}\omega)^{1-\eta}+c'_{\omega}\omega^{2}}.
\label{eq:AFMQCPSQUARE}
\end{align}
From the momentum dependence of the magnetic susceptibility, as shown in Fig.~\ref{fig:fig10}, we obtain a finite anomalous dimension $\eta=0.125$. 
For the frequency dependence, a crossover behavior is observed, and for quantity $\chi^{-1}\equiv \chi^{-1}(T,h_c,0,\omega)-\chi^{-1}(T,h_c,0,0)$, it is well described by
\begin{equation}
\chi^{-1}(\omega)=(c_{\omega}\omega)^{1-\eta}+ c'_\omega \omega^{2}.
\end{equation}
At low frequency, the first term with power law form $\omega^{0.875}$ dominant, consistent with the anomalous dimension $\eta=0.125$ found in momentum dependence. At higher frequency, the $\omega^2$-term dominants, back to the bare  bosonic susceptibility of (2+1)D Ising universality class. We want to remark here that although our QMC data is consistent with dynamical exponent $z=2$ of HMM, significant derivations in the form of anomalous dimension in the dynamical exponent as predicted in Ref.~\cite{Metlitski2010b} cannot be excluded. Interestingly, the $O(2)$ itinerant AFM-QCP case is considered in another recent numeric work~\cite{Gerlach2017}, where the dynamical exponent $z=2$ is found as well, consistent with the one-loop RG results, but higher order correction indeed exists in the latest RG calculation~\cite{Schlief2017}. However, for the Ising case we considered here, such similar calculations are still absent. It will be of great theoretical and numerical interests, that such a higher order RG calculation, can be performed in the AFM-QCP with Ising symmetry and $2\mathbf{Q}=\Gamma$ case as well.

\begin{figure}[htp!]
\centering
\includegraphics[width=\columnwidth]{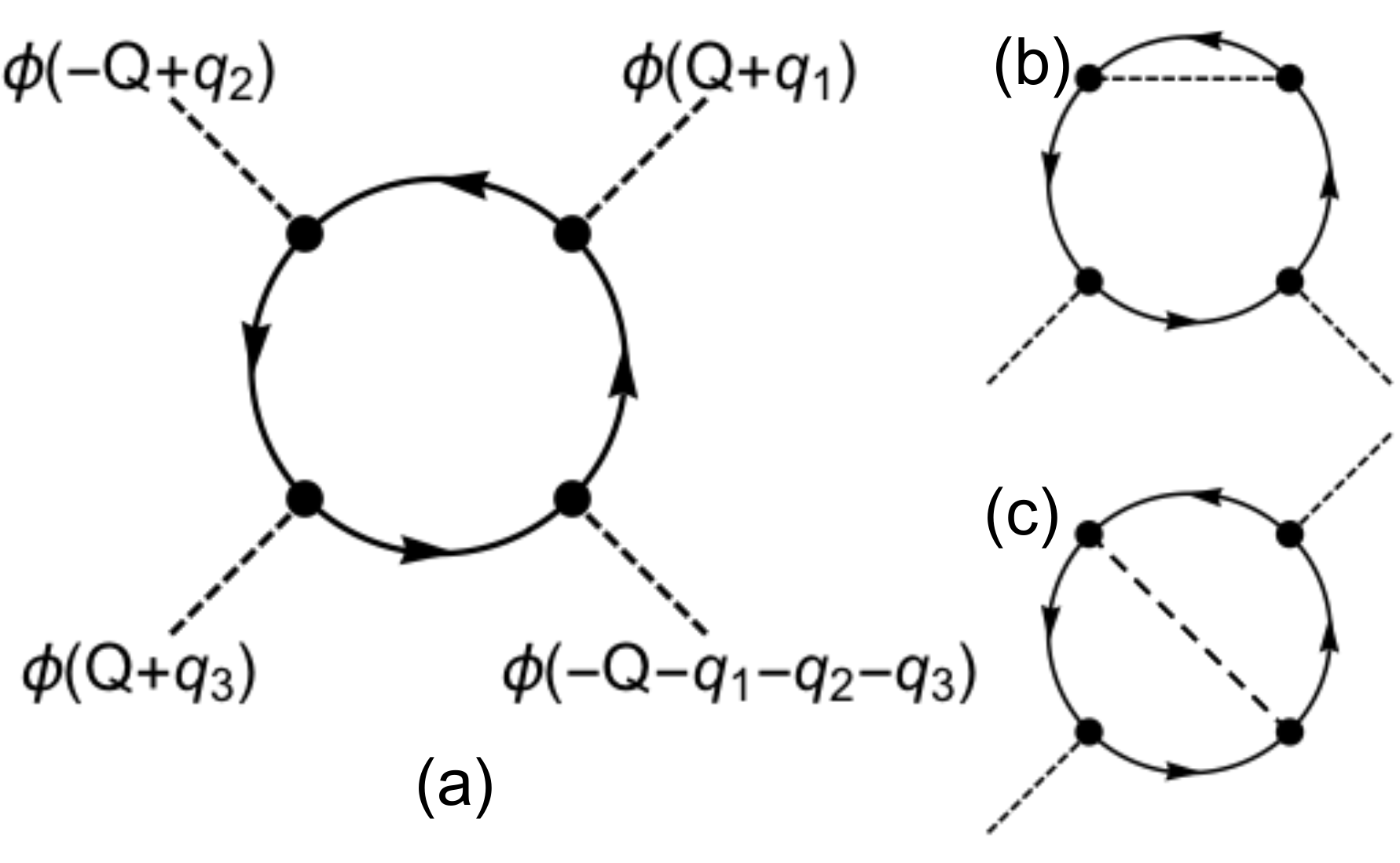}
\caption{(a) Feynman diagram representing a four-boson interaction vertex. Dashed lines, $\phi(\mathbf{k})$, represent spin fluctuations at momentum $\mathbf{k}$ and we set $q<<Q$. Because low-energy physics is dominated by fermionic excitations near the FS, two of the four boson legs must have momenta near $\mathbf{Q}$, while the other two are near $-\mathbf{Q}$ to keep the fermions near the FS as shown in the figure. For $2\mathbf{Q}=\Gamma$, $+\mathbf{Q}$ and $-\mathbf{Q}$ becomes identical, and thus there exist two ways to contract the external legs as shown in (b) and (c). For $2\mathbf{Q} \ne \Gamma$, however, only the contraction shown in (b) is allowed, while the momentum conservation law is violated in (c). Figures adapted from Ref.~\cite{ZiHongLiuSqu2018}.}
\label{fig:fig11}
\end{figure}

The anomalous dimension $\eta$ found in AFM-QCP with $2\mathbf{Q}=\Gamma$ case but not in AFM-QCP with $3\mathbf{Q}=\Gamma$ case is highly nontrivial and worth to be further discussed. Below we try to understand such difference and the implication of the presence of $\eta$ in the $2\mathbf{Q}=\Gamma$ case in more detail. 

First, between $2\mathbf{Q}=\Gamma$ and $3\mathbf{Q}=\Gamma$, the constraints dictated by the  momentum conservation law are different. The QCPs with $2\mathbf{Q}=\Gamma$ deviates from the HMM already at the level of four-boson vertex correction~\cite{Abanov2003}, as shown in Fig.~\ref{fig:fig11} (a). This four-boson vertex gives two topologically different bosonic self-energy diagrams as shown in Fig.~\ref{fig:fig11} (b) and (c) when $2\mathbf{Q}=\Gamma$, and particularly the diagram shown in Fig.~\ref{fig:fig11} (c) results in logarithmic corrections and is responsible for the breakdown of the HMM scaling. However for the $3\mathbf{Q} = \Gamma$ case in the triangular lattice model, this crucial diagram is prohibited by the momentum conservation law, hence the deviations from the HMM is not expected, at least within the same level of approximation. 

Second, let us discuss the meaning of the particular value of the anomalous dimension we found. Comparing to existence theories, the Heisenberg AFM-QCPs with $SU(2)$ symmetry has been considered in the literature~\cite{Abanov2003,Metlitski2010b}, while the Ising spin case is not yet carefully analyzed. We compare our numerical results with existing theoretical predictions from Heisenberg AFM-QCPs, qualitative but not quantitative agreement is expected because of the symmetry difference. In the calculations in Refs~\cite{Abanov2003,Metlitski2010b}, the anomalous dimension depends on the angle between Fermi velocity at the hot spots and the AF wavevector $\mathbf{Q}$ of order parameter. In our model, we have the angle close to $45^\circ$. But the RG calculations in Refs.~\cite{Abanov2003,Metlitski2010b} actually assume the angle equals zero degree at the fixed point, and only when that happens, the RG predicted $\eta=1/N_{h.s.}$ where $N_{h.s.}$ is the number of hot spots. In our model, we have $N_{h.s.}=16$, thus RG predicted $\eta=1/16$, while our QMC results find qualitatively consistency but with a value close to $2/N_{h.s.}$. The reason of the difference, possible comes from the symmetry difference or any other contributions, is an open issue, and further RG calculations on the AFM-QCP with Ising symmetry is highly desirable.

Another prediction from RG calculation is that near the QCP, the Fermi surface at hot spots will rotate towards nesting~\cite{Abanov2003}, and it further increases the anomalous dimension and can even renormalize dynamical exponent $z$~\cite{Abanov2003,Metlitski2010b}. Such a rotation of Fermi surface is indeed observed in very recent QMC studies~\cite{ZiHongLiuSqu2018}. However, the rotation we observed is small due to very slow RG flow and thus the resulting changes in anomalous dimension and dynamical exponent is beyond numerical resolution at the current stage. Even larger system sizes and lower temperatures, or better designer Hamiltonians, are needed to be accessed in future QMC simulations.

\begin{figure}
	\includegraphics[width=\columnwidth]{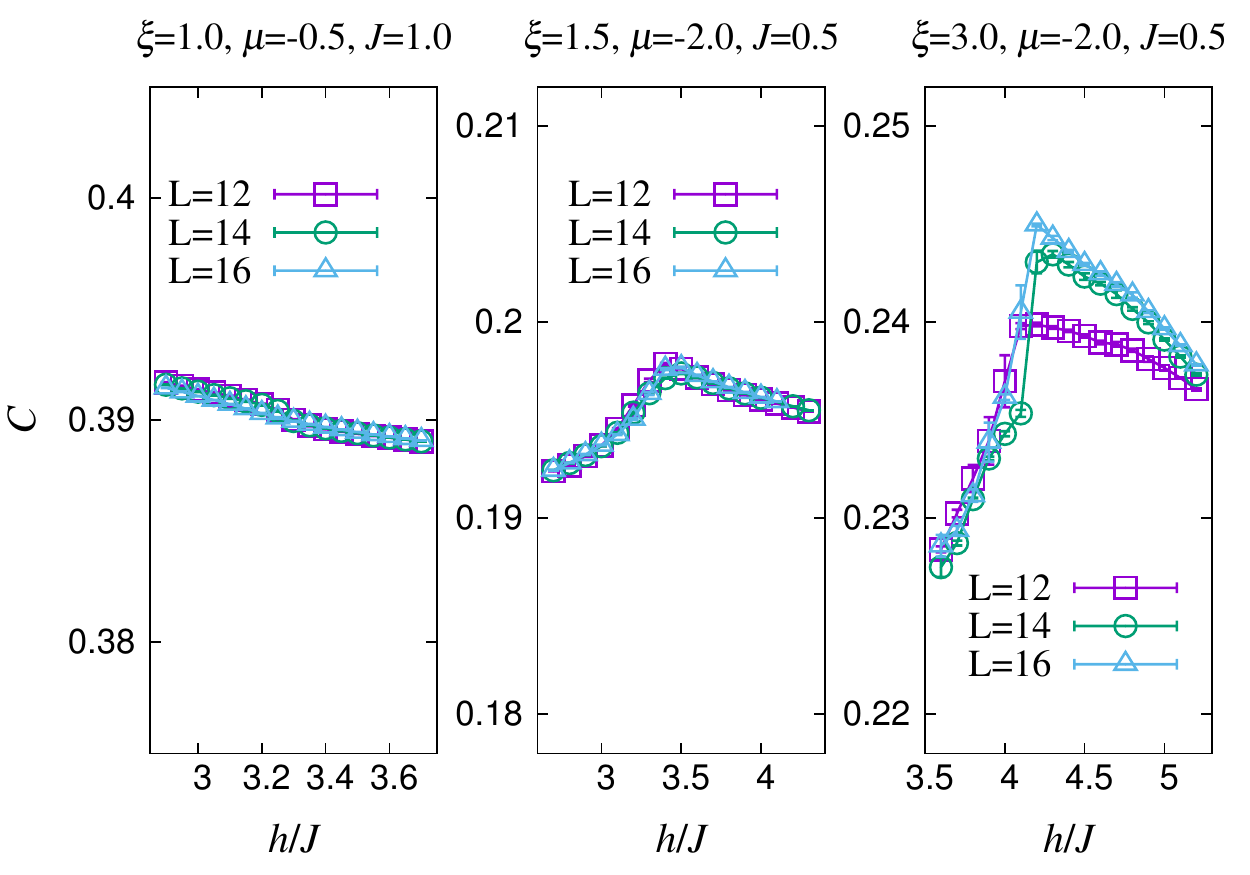}
	\caption{Static pairing-correlation function $C=\frac{1}{L^{2}}\langle\hat{\Delta}^{\dagger}\hat{\Delta}\rangle$ for order parameters defined in Eq.(11) in Ref.~\cite{Xu2017}. For \{$\xi=1.0,J=1.0,\mu=-0.5$\}, no enhancement of pairing correlation functions is observed in any pairing channel down to $T=0.025$. For \{$\xi=1.5,J=0.5,\mu=-2$\} and \{$\xi=3.0,J=0.5,\mu=-2$\}, the pairing order parameters $\Delta_\uparrow$ and $\Delta_\downarrow$ show enhanced correlation near the QCP, in agreement with theoretical analysis. No enhancement is observed in other pairing channels. Figures are adapted from Ref.~\cite{Xu2017}.}
	\label{fig:fig12}
\end{figure}

\subsection{Superconductivity}
In the vicinity of QCP, other instabilities like superconductivity may emerge, which usually may bury the QCP, pollute QCR and prevent us to do a faithful scaling analysis down to very low temperature. The construction we make seems to enjoy a very pristine QCP that does not develop any superconductivity down to lowest temperature we studied with coupling constant $\xi=1$. Further increasing $\xi$, it may finally drives a superconductivity instabilities, which is also find in other related numerical studies, such as in Refs.~\cite{Schattner2015b,Lederer2016,Gerlach2017}.

Fig.~\ref{fig:fig12} shows the superconductivity correlations for different coupling constant in FM-QCP case. We do find as the coupling constant $\xi$ increases, the pairing correlations start to increase with system size near QCP, which is a signature of pairing instability.

\section{Quantum criticalities of fermions coupled to phonons, $\mathbb{Z}_2$ or $U(1)$ gauge fields}
\label{sec:gaugeQCP}

The philosophy of designer Hamiltonian in Eq.~\eqref{eq:generalformula} is generic, besides couple fermion to critical bosonic modes in the forms of ferromagnetic, antiferromagnetic and nematic fluctuations, the boson modes can be extended to other situations such as Holstein photons~\cite{Johnston2013,ChuangChen2018a,ChuangChen2018b} to study the electron-phonon interaction mediated metal to charge density wave (CDW) transition~\cite{Johnston2013,ChuangChen2018a,ChuangChen2018b} and metal to superconductivity transition~\cite{Esterlis2018,Costa2018,ZXLi2018}, and these interaction-driven phases. Along this line of research, new quantum Monte scheme in the form of SLMC which introduces symmetry-enforced global update of phonon field to reduce the autocorrelation time~\cite{ChuangChen2018a} and Langevin dynamics~\cite{BATROUNI1987,Batrouni2019} which performs efficient global update to reduce the computational complexity have been successfully introduced. The situation, that simulating Holstein model at 2D or higher was extremely difficult compared with other DQMC simulations for, say, Hubbard model~\cite{Hohenadler2008}, has been profoundly changed in the last few years. 

More interestingly, the exact form of critical bosons can even go beyond the conventional condensed matter wisdom and acquire incarnations with high-energy flavor, such that they could play the role of gauge degree of freedom and carry gauge symmetry and topological orders. In this way, the situation of matter fields couples to gauge fields can be realized in our setting and the rich physics of fractionalization of electrons to anyons coupled to the emergent gauge fields of $\mathbb{Z}_2$~\cite{Assaad2016,Gazit2016,Gazit2018,ChuangChen2019,Gazit2019} and $U(1)$~\cite{XiaoYanXu2018U1,WeiWang2019} symmetries in many strongly correlated systems such as $\mathbb{Z}_2$ and $U(1)$ topological orders~\cite{WenZ2SL1991,Senthil2000,Rantner2002,Hermele2005} and their material and model realizations in quantum spin liquids~\cite{Fu2015,Feng2017,hermele2004a,GYSun2018,YCWang2018,Huang2018} and deconfined quantum criticalities~\cite{Senthil2004,Sandvik2007,Qin2017,Ma2018}, and the fundamental question of the existence of deconfinement at (2+1)D quantum electrodynamics (QED$_3$) with fermionic matter in high-energy physics~\cite{Polyakov1977,burkitt1988glueballs,Fiebig1990,Armour2011}, can now be addressed with unbiased quantum Monte Carlo simulations. The development in this direction is fast and profound. 

In this section, we try to briefly summarize the activities in the aforementioned areas in recent years, based primarily upon the works of our own.

\subsection{Dirac fermions coupled with phonons}
As mentioned above, although the electron-phonon coupled systems, realized in the Holstein model or Holstein-Hubbard model, play important role in the understanding of metal to charge density wave (CDW) transition~\cite{Johnston2013,ChuangChen2018a,ChuangChen2018b,YXZhang2019} and metal to superconductor transition~\cite{Esterlis2018,Costa2018,ZXLi2018} and these interaction-driven phases themselves, the quantum Monte Carlo simulations of Holstein model in 2D or higher dimensions are notoriously difficult due to strong autocorrelation times~\cite{Hohenadler2008}. The continuous phonon fields render the local update of conventional DQMC inefficient and the growth of autocorrelation time in the DQMC of Holstein model can be shown much faster than that of the Hubbard model, see Fig.2 (a) in Ref.~\cite{ChuangChen2018a} of 2D Holstein model at metal-to-CDW transition temperature for example, the measured autocorrelation time $\tau \sim L^{5}$, much larger than the typical $\tau \sim L^{2}$ for the local update of Metropolis algorithms. Recently, we applied SLMC method on 2D Holstein model and found an efficient way of designing the effective bosonic Hamiltonian that enforces the $Z_2$ symmetry of the on-site energy potential of the phonons, in this way, the SLMC of Holstein model can reduce the autocorrelation time to $\tau \sim L^{2}$~\cite{ChuangChen2018a} and makes the 2D lattice systems simulatable.

With such development, we investigated the electron-phonon coupled problem on 2D honeycomb lattice~\cite{ChuangChen2018b}. This model is relevant to the electron-phonon coupling in graphene and the twisted angle graphene multilayers with superlattices~\cite{YuanCao2019,cao2018correlated,ChengShen2019}. The Dirac fermion in the graphene honeycomb lattice is robust against the weak coupling with phonons. But when the coupling is strong enough, the fluctuations of the phonons modes would generate effective attraction between electrons and eventually lead to a CDW insulator. In the work of Ref.~\cite{ChuangChen2018b}, such a metal-to-CDW transition is studied with unprecedentedly large system sizes (qauntum phase transition at $T=0$ with electron number of $2\times L\times L$ upto $2\times 15\times 15$), and the transition is revealed with dynamically generated mass of Dirac fermion and belongs to $N=2$ flavors (2+1)D chiral Gross-Neveu universality~\cite{YYHe2018,Chandrasekharan2013,Zerf2017}. The critical exponents are determined and consistent with previous quantum Monte Carlo results with the same universality but physically different systems~\cite{YYHe2018,Chandrasekharan2013}. The comparison of the exponents with high order perturbative RG calculation~\cite{Zerf2017} is still not completely satisfactory and this would require future works with even larger system sizes, also more controlled calculation from the theoretical end. Judging from the developments and achievements summarized in this review, we are confident that eventually with the mutual stimulation and developments from numerical and analytical communities, the inconsistency in the chiral Gross-Neveu universities of the gapped out Dirac fermions with spontaneous fermion bilinear of symmetry-breaking, would be cured.


\subsection{Dirac fermions coupled with $\mathbb{Z}_2$ or $U(1)$ gauge fileds}
Also as aforementioned, in the other direction where fermions are coupled to $\mathbb{Z}_2$ or $U(1)$ gauge fields, the system becomes directly related to both high energy and condensed matter physics. This is because in the high energy physics, there exists a fundamental question of whether the deconfined phase exists or not in the setting of fermion coupled to the gauge field in QED$_3$, where it is known that the pure gauge field in QED$_3$ will always be confined due to monopole proliferation~\cite{Polyakov1977}, but with the help of the matter fields, the situation might be different and this question relies on the unbiased quantum Monte Carlo simulation to verify. And in the condensed matter physics setting, the situation of the fermions couple to gauge fields is believed to provide the low energy descriptions of many condensed matter systems ranging from  high-temperature superconductors~\cite{PatrickLee1998,PatrickLee2006} to $\mathbb{Z}_2$ or $U(1)$ quantum spin liquids~\cite{Fu2015,Feng2017,hermele2004a,GYSun2018,YCWang2018,Huang2018} and deconfined quantum critical points~\cite{Senthil2004,Sandvik2007,Qin2017,Ma2018}, etc. Due to these reasons, the quantum Monte studies of matter fields coupled to gauge fields are of great interests to very broad audience.

The Dirac fermions coupled with $\mathbb{Z}_2$ gauge fields in 2D systems are studied in Refs.~\cite{Gazit2016,Assaad2016,Gazit2018,Gazit2019}. In Refs.~\cite{Gazit2016,Gazit2018}, the gauge constrain is enforced in the Monte Carlo simulation and in Ref.~\cite{Assaad2016}, the gauge constrain is dynamically generated at low temperature. But overall, similarly rich phase diagrams are established at different number of fermion flavors and the strength of the coupling of gauge field with matter field (Dirac fermions). There always exists a deconfined phase where the Dirac cones in the matter fields persist and coexist with an emergent $\mathbb{Z}_2$ topological order. As the coupling becomes weak, the systems enter a confined phase via deconfinement-to-confinement phase transition, and depending on the number of fermion flavors, the confined phases develop various symmetry-breaking in the form of CDW, valence bond solid (VBS) and AFM phases, in all these phases, the Dirac fermion is gapped out. In the work of Ref.~\cite{Gazit2018}, a on-site Hubbard $U$ term is added into the Hamiltonian, and the large $U$ limit is also a AFM phase but with $\mathbb{Z}_2$ topological order. Therefore, there is another deconfinement-to-confined phase transition between two different AFM phases. The transition from deconfined phase to AFM with confinement is found to possibly host an emergent $SO(5)$ symmetry~\cite{Gazit2018}. All these results are of great interests. And the question of Fermi surface coupled to $\mathbb{Z}_2$ topological order via the help of $\mathbb{Z}_2$ matter field has recently been address by some of us~\cite{ChuangChen2019}.

More recently, Dirac fermions coupled $U(1)$ gauge fields problem is studied by some of us~\cite{XiaoYanXu2018U1,WeiWang2019}. Our designer Hamiltonian, in the spirit of Eq.~\eqref{eq:generalformula}, is a 2D quantum rotor model which couples Dirac fermions with compact $U(1)$ gauge field on 2D square lattice,
\begin{eqnarray}
H&=&\frac{1}{2}JN_{f}\sum_{\langle i,j \rangle} \frac 1 4 \hat{L}^{2}_{ij}-t\sum_{\langle i,j \rangle\alpha}\left(\hat{c}^{\dagger}_{i\alpha}e^{i\hat{\theta}_{ij}}\hat{c}_{j\alpha}+\text{h.c.}\right) \nonumber\\
&& +\ \frac{1}{2}K\ N_f\sum_{\square}\cos \left( \text{curl} \hat{\theta} \right),
\label{eq:u1hamiltonian}
\end{eqnarray}
where $N_f$ is the number of fermion flavors, $\hat{L}_{ij}$ are canonical angular momentum and $\hat{\theta}_{ij}$ are coordinate operators, and they satisfy relation $[\hat{L}_{ij},e^{\pm i \hat{\theta}_{ij}}] =
\pm e^{\pm i\hat{\theta}_{ij}}$. The Hamiltonian defined in Eq.~\eqref{eq:u1hamiltonian} can be formulated in a path integral with action
\begin{equation}
S=S_F+S_{\phi}=\int_0^{\beta}d\tau (L_F+L_\phi)
\end{equation}
where
\begin{eqnarray}
L_F &=& \sum_{\langle ij \rangle\alpha}{\psi}^{\dagger}_{i\alpha} \left[(\partial_\tau -\mu)\delta_{ij}-t e^{i\phi_{ij}}   \right]   {\psi}_{j\alpha} + \text{h.c.}, \nonumber \\
L_\phi &=& \frac{4} {JN_{f}\Delta \tau ^2} \sum_{\langle ij \rangle}
\left( 1-\cos(\phi_{ij}(\tau+1)-\phi_{ij}(\tau)) \right) \nonumber \\
&& +\frac{1}{2}K N_f\sum_{\square}\cos \left( \text{curl} \phi \right).
\label{eq:lagrangian}
\end{eqnarray}

\begin{figure}[t]
\includegraphics[width=\columnwidth]{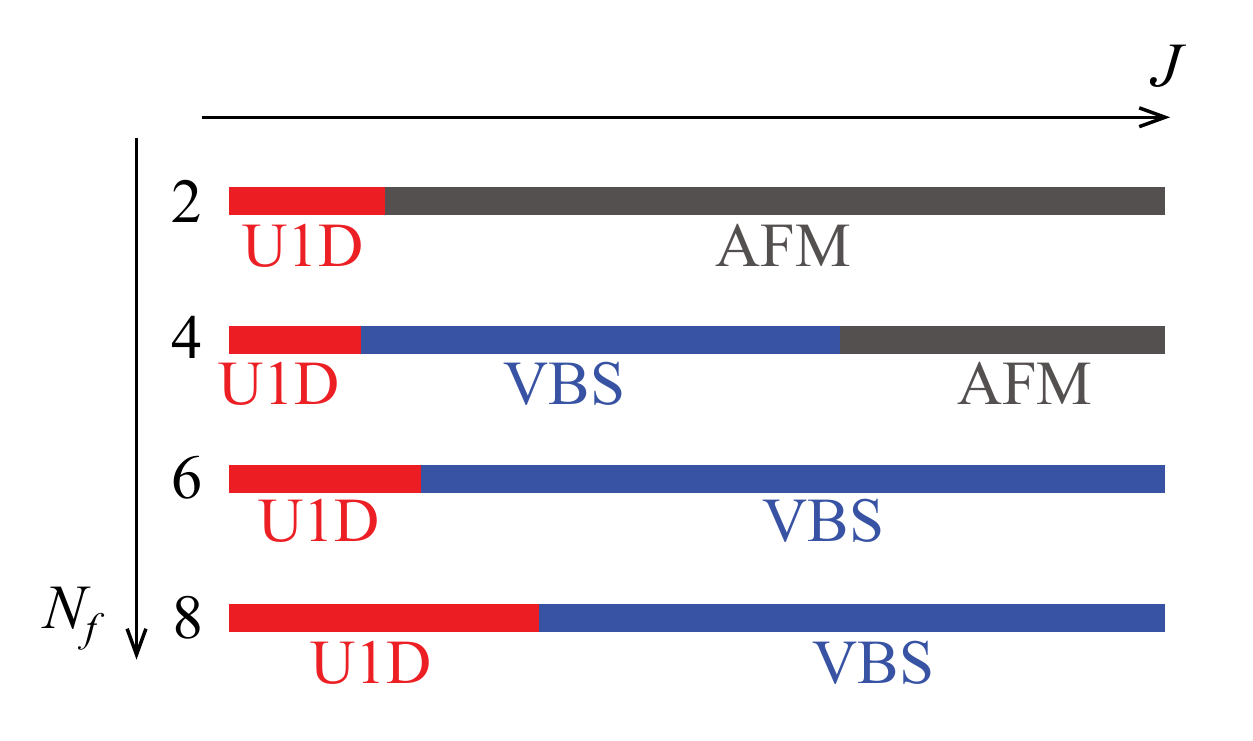}
\caption{Phase diagram spanned by the fermion flavors $N_f$ and the strength of gauge field fluctuations $J$ of the model shown in the Hamiltonian Eq.~\eqref{eq:u1hamiltonian}. U1D stands for the $U_1$ deconfined phase where the fermions dynamically form a Dirac system. This phase corresponds to the algebraic spin liquid~\cite{Hermele2005} where all correlation functions show slow power-law decay. VBS stands for valence bond solid phase and AFM stands for the antiferromagnetic
long-range ordered phase. The figure is adapted from Ref.~\cite{XiaoYanXu2018U1}.}
\label{fig:fig13}
\end{figure}

This model is sign problem free when number of fermion flavors $N_f$  is even due to pseudo-unitary symmetry which guarantee each determinant is a real number (see Appendix A. in Ref.~\cite{XiaoYanXu2018U1} for the rigorous proof). We focus on $K=t=1$ and explore the phase diagram when turning $J$ for different $N_f$ as showed in Fig.~\ref{fig:fig13}. We studied cases when $N_f=2$, 4, 6 and 8 respectively, and discovered the $U(1)$ deconfined phase (U1D) and confinement transition in each case. The properties of U1D phase are consistent with the proposal of algebraic spin liquid, where various competing orders (such as AFM order, VBS order etc.) all have identical power-laws algebraic correlation in real space.  Importantly, we found the decaying power match perfectly with the larger-$N_f$ perturbative renormalization group expression~\cite{Rantner2002,Hermele2005,Hermele2007,Xu2007}. 
The continuous confined transitions from U1D to AFM or VBS we found should be described by QED$_3$-Gross-Neveu O(2) or O(3) universality, depending on the symmetry group that the fermion bilinears break in the confined phase, and further carefully study of the critical properties of these transitions via QMC simulations and analytical calculations is certainly worthwhile~\cite{WeiWang2019}. Recently perturbative renormalization group calculations to higher orders have been carried out in attempt to accquire the critical properties of the deconfinement to confinement transition in form of QED$_3$-Gross-Neveu universality classes~\cite{Janssen2017,Ihrig2018,Nikolai2018,Gracey2018,Boyack2019,Zerf2019} and to address the relevance or irrelevance of the monopole operators to the stability of the U1D phase~\cite{XYSong2018a,XYSong2018b,Dupuis2019}, show substantial interests and great ongoing efforts along this direction.

The discovery of the stable U1D phase at low fermion flavor numbers (possibly starting from $N_f=2$) is of vital importance as the situation corresponds to the experimental relevant case. Further more, the phase diagram is of importance, and we believe that it will trigger a number of future investigations on the numerical and analytical fronts, both in the solid state and the high energy communities.

\section{Discussion and prospectives}
In this review, we summarize the recent progress in developing and employing unbiased quantum Monte Carlo techniques, to investigate the fermion quantum criticalities that can be realized in the designer models such as Eqs.~\eqref{eq:model}, ~\eqref{eq:generalformula} and ~\eqref{eq:specificformula}. In earlier parts of this review, we provided a highly condensed summary about the long history and the intensive theoretical efforts that have been devoted to understand these highly interesting quantum systems, as well as key challenges and difficulties that hold us from obtaining full and final solutions. Nevertheless, great theoretical progress has been made since the HMM framework, and new concepts in the RG and in nFL, hot spots, anomalous transport and quantum critical scaling behaviors prevail. 

On the other hand, it took the numerical community quite some time to understand and appreciate these modern theoretical developments and finally is able to incorporate them into the developments of numerical methodologies, exemplified here in the large-scale quantum Monte Carlo simulations. Instead of the treating explicit four-fermion interactions like the Hubbard interaction, which was the focus of the numerical community since the early date of DQMC, in the new designer models discussed in this review, one can directly couple the FS to various critical bosonic modes driven by the consideration of the physical question one would like address, ferromagnetic, antiferromagnetic, nematic critical bosonic fluctuations and QCPs, just to name a few. In short, one can design Hamiltonians and to directly simulate the situations that have been asked in the field theoretical analyses/hypotheses mentioned above. Such awareness has greatly liberated the mindset of the numerical community and new models and results from QMC simulations thus flourishes. Along this process, better numerical methods are also invented from the mutual inspiration and dialogues between numerical and theoretical communities, SLMC, EMUS are the successfully examples.

The works presented here, contain the fermionic QCP in 2D systems on square, triangle lattices, with the bosonic fluctuations in ferromagnetic, antiferromagnetic forms acquired Ising symmetry. We found that in the FM-QCP case, the entire FS becomes critical and the quasiparticle vanishes at the critical FS, the quantum critical scaling clearly acquires finite anomalous dimension different from the 3D Ising universality and that of the HMM prediction, the Lorentz symmetry between the space and time is also lost at the FM-QCP. All this evidence is pointing to a new and unknown universality. Our results provide the valuable references for the further development in the field theoretical treatment to this difficult problem. 

As for the AFM-QCPs, we found an interesting difference between the $2\mathbf{Q}=\Gamma$ and $3\mathbf{Q}=\Gamma$ situations, where the former is also clearly deviated from the HMM and acquires the anomalous dimension, the latter is actually consistent with the HMM prediction which enhances the dynamical exponent from $z=1$ in the bare boson case to the $z=2$ in the coupled case. The $2\mathbf{Q}=\Gamma$ is unique in the sense that this is for the first time, an finite anomalous dimension with $\eta=0.125$ is discovered in the unbiased numerical simulations and shining light on the correctness in the previous RG calculation that the $\eta$ might be obtained from perturbative calculation as a function of $N_{\text{h.s.}}$. However, the discrepancy still exists and this might due to the fact that the existing theoretical calculations are performed with $O(3)$ bosons whereas our results are for $Z_2$ bosons. Nevertheless, it is clear that from here, both theoretical and numerical communities can adjust their models and try to perform the first exact comparison with exactly the same low-energy physical condition and to see whether the same fixed point can be reached.

Above are the main content of the review, and in Sec.~\ref{sec:gaugeQCP}, we also summarized the recent activities in other related directions such as the fermions coupled to phonons and fermions coupled to gauge degree of freedoms. The progresses in these directions, especially the later ones, go even beyond the boundary of the condensed matter physics and are bridging out to the fundamental question such as the existence of deconfinement in the QED$_3$ in high energy physics. The existence of the $\mathbb{Z}_2$ and $U(1)$ gauge field theory with fermion modes provide the concrete examples of the unconventional metallic phases which might stem from the fractionalization of electrons into anyons with associated emergent gauge fields in many strongly correlated systems that are the breeding ground of the quantum state of matter beyond conventional paradigms of FL and spontaneous symmetry breaking. Examples include topological ordered states and quantum spin liquids, Cuprate high-temperature superconductors, etc.

In addition to results summarized above, this progress in model design and numerical techniques offers a platform to explore other challenging problems in strongly correlated systems. In particular, this platform finally enables direct comparison between unbiased numerical simulations and theoretical predictions, and thus a closed feedback loop between theory and numerical studies can be formed. We can now utilize unbiased numerical results to verify and refine our theoretical knowledge, and the progress on the theory side will in return provide more guidance for numerical exploration. For classical critical phenomena, such a feedback loop has played a vital role since decades ago and has resulted in highly fruitful results and achievements. In quantum systems, we have good reasons to believe that through the same type of theory-numerical iterations, we are getting closer than ever before to the fully understanding of strongly correlated phenomena, such as fermionic quantum criticality and beyond.
From a humble beginning, an odyssey of discovery has lead us to this stage, that based on the efforts mentioned in this review, the numerical community and theoretical one can finally sit down and start to compare each others results. From this point, one could expect the solution of the fermionic QCPs, or, put in the better terms, the concrete framework of the fermionic QCPs beyond that of the HMM, as concrete as that of the Landau-Ginzburg or Wilson-Fisher, are expected to be reached in the near future.

\begin{acknowledgements}
We would like to express our gratitude towards Fakher Assaad, Yoni Schattner, Erez Berg, Richard Scalettar, George Batrouni, Martin Hohenadler, Simon Trebst and Snir Gazit upon the illustrative discussions and kind support from the numeric perspectives throughout the simulations of the works presented in this review. Likewise, we would like to acknowledge Andrey Chubukov, Subir Sachdev, Steven Kivelson, Max Metlitski, Sung-Sik Lee and Michael Scherer on the very helpful discussions and education on theoretical understanding of modern developments of the itinerant quantum critcalities. XYX, ZHL, GPP and ZYM acknowledge the supports from the Ministry of Science and Technology of China through the National Key Research and Development Program (2016YFA0300502), the Strategic Priority Research Program
of the Chinese Academy of Sciences (XDB28000000), and the National Science Foundation of China (11421092,11574359,11674370). XYX is also thankful for the support of HKRGC through grant C6026-16W. YQ acknowledges support from Minstry of Science and Technology of China under grant numbers 2015CB921700, and from National Science Foundation of China under grant number 11874115.   K.S. acknowledges support from the National Science Foundation under Grant No. EFRI-1741618 and the Alfred P. Sloan Foundation. We thank the Center for Quantum Simulation Sciences in the Institute of Physics, Chinese Academy of Sciences, the Tianhe-1A platform at the National Supercomputer Center in Tianjin and Tianhe-2 platform at the National Supercomputer Center in Guangzhou for their technical support and generous allocation of CPU time.
\end{acknowledgements}

\bibliography{main}

\end{document}